\documentclass[traditabstract]{aa}
\usepackage{graphicx,natbib}

\newcommand{\gx}{GX~339-4\ \ignorespaces}

\def\lta{ \lower .75ex\hbox{$\sim$} \llap{\raise .27ex \hbox{$<$}} }

\begin{document}

\date{Received .../Accepted ...}

\title{The return to the hard state of GX 339-4 as seen by Suzaku}
%disk magnetization: a crucial parameter in the hysteresis cycle of X-ray binaries}
\author{P.-O. Petrucci\inst{1}
  \and C. Cabanac\inst{2}
  \and S. Corbel\inst{3}
  \and E. Koerding\inst{4}
  \and R. Fender\inst{5,6}
  }
  
\institute{$^1$ UJF-Grenoble 1 / CNRS-INSU, Institut de Plan\'etologie et d'Astrophysique
	   de Grenoble (IPAG) UMR 5274, Grenoble, F-38041, France\\
	   $^2$ Universit\'e de Toulouse, UPS-OMP, IRAP, Toulouse, France CNRS, IRAP, 9, av du Colonel Roche, BP 44346, F-31028 Toulouse Cedex 4, France\\
	   $^3$ Laboratoire AIM (CEA/IRFU - CNRS/INSU - Universit\'e Paris Diderot), CEA DSM/IRFU/SAp, F-91191 Gif-sur-Yvette, France\\
	   $^4$ Department of Astrophysics/IMAPP, Radboud University Nijmegen, P.O. Box 9010;6500 GL Nijmegen;The Netherlands\\
	   $^5$ University of Oxford, Department of Physics, Astrophysics, Denys Wilkinson Building, Keble Road, OX1 3RH, Oxford, UK\\
	   $^6$ School of Physics and Astronomy, University of Southampton, Highfield, Southampton, SO17 1BJ, UK	
          } 

%\author{P.O.\,Petrucci\inst{1} et al.}
%
%\institute{UJF-Grenoble 1 / CNRS-INSU, Institut de PlanŽtologie et d'Astrophysique de Grenoble (IPAG) UMR 5274, Grenoble, F-38041, France }

\abstract{The reality of the disk recession is an information of prime importance to understand the physics of the state transitions in X-ray binaries. The microquasar GX 339-4 was observed by Suzaku five times, spaced by a few days, during its transition back to the hard state at the end of its 2010-2011 outburst. The 2-10 keV source flux decreases by a factor $\sim$ 10 between the beginning and the end of the monitoring. Simultaneous radio and OIR observations highlighted the re-ignition of the radio emission just before the beginning of the campaign, the maximum radio emission being reached between the two first Suzaku pointings, while the IR peaked a few weeks latter.  A fluorescent iron line is always significantly detected. %, and is consistent with neutral iron in all observations but OBS3. 
Fits with a gaussian or Laor profiles give statistically equivalent results. In the case of a Laor profile, fits of the five data sets simultaneously agree with a disk inclination angle of $\sim$20 degrees. The disk inner radius is $<10-30\ R_g$ in the first two observations but almost unconstrained in the last three due to the lower statistics. A soft X-ray excess is also present in these two first observations. Fits with a multicolor disk component give disk inner radii in qualitative agreement with those obtained with the iron line fits. The use of a physically more realistic model, including a blurred reflection component and a comptonization continuum, give some hints of the increase of the disk inner radius but the significances are always weak (and model dependent) preventing any clear conclusion concerning disk recession during this campaign. %On the other hand, the disk reflection properties appear clearly different between the two first observations and  the last three ones, with a clear increase of the surface layer hydrogen surface density and of  the illuminating to intrinsic disk flux ratio. This evolution agrees with the OIR reflare peaking between OBS2 and 3, and which may reflect the necessary time to build up enough density in the jet. The observed changes of the disk properties could correspond to the natural evolution from a standard (i.e. non ejecting) disk to a jet emitting disk. 
Interestingly, the addition of warm absorption significantly improves the fit of OBS1 while it is not needed in the other observations. The radio-jet re-ignition occurring between OBS1 and OBS2, these absorption features may indicate the natural evolution of the accretion outflows transiting from a disk wind, an ubiquitous characteristic of soft states, and a jet, signature of the hard states. The comparison with a long 2008 Suzaku observation of GX 339-4 in a persistent faint hard state (similar in flux to OBS5) where a narrow iron line clearly indicates a disk recession, is discussed.  }

%Black hole physics -- Accretion, accretion discs -- Magnetohydrodynamics (MHD) -- ISM: jets and outflows -- X-rays: binaries}

\maketitle

\section{Introduction}
Multi-wavelength observation campaigns of microquasars, like those done in X-ray and Radio in the last 10 years, were crucial to bring to light the strikingly link between the ejection phenomena (mostly observed in radio) and the inner accretion flow, whose radiation seems to be the dominant component in the X-rays (e.g. \citealt{cor00,cor03,cori11,gal03}). %(Corbel et al. 2000, 2003, 2011; Gallo et al. 2003).  
For instance, strong radio emission, interpreted by the presence of persistent jets (directly observed in a few cases, e.g. \citealt{dha00,stir01}), is generally detected when the X-ray emission peaks at a few tens of keV, in the so-called hard state (e.g. \citealt{cor04,fen04}). This X-ray emission is commonly believed to originate via inverse Compton process from a plasma of hot electrons (the so-called corona) scattering off UV/soft X-ray photons produced by the cooler part of the accretion flow. On the other hand, in the so-called soft state the radio emission is quenched \citep{fen99,cor00} and the X-ray data are spectrally dominated by soft X-ray emission. This emission is generally interpreted as signature of a multi-color accretion disk component down to the last stable orbit $R_{ISCO}$. 

{In the past ten years, high energy resolution observations of several microquasars showed also the presence of highly ionized absorption features in their X-ray spectra. These features were interpreted as signature of ionized gaz in the close environment of the black hole, their blueshifts being indication of outflows or winds (e.g. \citealt{mil04a} hereafter M04, \citealt{mil06c}). It has been realized that these features were more specifically observed in the soft state (e.g. \citealt{pon12,dia11,dia12b}). These results suggest that, during outbursts, X-ray binaries may transit back and forth between disk-jet (in the hard state) and disk-wind (in the soft state) configurations (e.g. \citealt{nei09}}).  However the exact interplay between ejection and accretion phenomena and the origin of the transition from one state to the others is still poorly known (see however the recent study done by \citealt{kal13}).

%Microquasars are known to evolve between different spectral states during their outbursts. At the beginning and the end of these outbursts, they are in the so-called hard states, characterized, above $\sim$3 keV, by power-law like spectral energy distribution (SED) peaking in the hard X-rays ($\sim$100 keV, e.g. \citealt{rem06,don07}) before a high energy cut-off. This emission is generally believed to originate via inverse compton emission from a plasma of hot electrons (the so-called corona) scattering off UV/soft X-ray photons produced by the accretion flow. When detected, the excess of flux above the power law extrapolation in the soft X-rays is commonly interpreted as the radiative signature of this accretion flow. Hard states are also characterized by strong radio emission, signature of powerful ejection.\\
%
%In the central part of the outburst, microquasars spectra are in the so-called soft states, the SEDs peaking in the soft X-rays and well fitted by a multi-color accretion disk component. A high energy tail (up to few MeV), well below the soft X-ray peak, can also be observed in this state, whose origin is still poorly understood. The radio emission is generally weak if not absent, and is interpreted by the ceasing of the ejection processes. In between these hard and soft states, the systems transit trough intermediate states characterized by complex spectral and timing properties \citep{rem06,don07}.\\
\begin{table*}
\begin{center}
\begin{tabular}{cccccccc}
Obs name & Obs ID & MJD & XIS03 Exp. & XIS 0-3 & XIS 0-3 & HXD/PIN& HXD/GSO\\ 
 & &  & ks & 0.7-2 keV ($s^{-1}$) & 2-10 keV ($s^{-1}$) & 20-70 keV ($s^{-1}$) & 70-200 keV ($s^{-1}$)\\ 
\hline
OBS1 & 405063010 & 55603.7 & 44.2 & 11.90$\pm$0.02 & 10.22$\pm$0.02 & 0.62$\pm$0.01 & 0.96$\pm$0.04\\
OBS2 & 405063020 & 55608.9 & 42.0 & 5.20$\pm$0.01 & 6.55$\pm$ 0.01& 0.46$\pm$0.01 & 1.05$\pm$0.04\\
OBS3 & 405063030 & 55616.8 & 38.4 & 1.81$\pm$0.01 & 2.80$\pm$ 0.01& 0.20$\pm$0.01 & 0.74$\pm$0.05\\
OBS4 & 405063040 & 55620.2 & 43.6 & 1.32$\pm$0.01 & 2.09$\pm$ 0.01& 0.15$\pm$0.01 & 0.74$\pm$0.04\\
OBS5 & 405063050 & 55627.5 & 37.3 & 0.92$\pm$0.01 & 1.50$\pm$ 0.01& 0.10$\pm$0.01 & 0.56$\pm$0.05\\
\hline
\end{tabular}\\
\caption{\label{tab1} Log of the 5 observations with their ID number, the corresponding date (in MJD), the sum of the XIS0 and XIS3 exposure time (each exposure being $\sim$20 ks, the total exposure of the XIS0 and XIS3 instruments is about twice this time) and the count rates in the different Suzaku instruments. .}
\end{center}
\end{table*}

The commonly adopted toy-picture of the central regions of microquasars plays on the relative importance of the accretion disk and hot corona emission along the outburst (e.g. \citealt{esi97,don07}). The accretion disk is assumed to be present in between an outer  and inner radius $R_{out}$ and $R_{in}$ while the hot corona is localized in between $R_{in}$ and $R_{ISCO}$. In the hard state, $R_{in}\gg R_{ISCO}$ and the hot corona dominates the observed emission. The inner part of the accretion disk (close to $R_{in}$) is then cold, explaining its poor detection in the soft X-rays in this state. Reversely, in the soft states $R_{in}\sim R_{ISCO}$, i.e. the hot corona is no more present and the spectra are dominated by the accretion disk emission.\\

If this picture is correct, variations of the disk inner radius $R_{in}$ should occur during the state transitions, first decreasing during the hard-to-soft transition but then increasing during the soft-to-hard one. This interpretation is apparently supported by the observations, in hard states, of weak reflection components (e.g. \citealt{gier97,bar00,mil02a,zyc98,joi07}), the absence of relativistic broadening of the iron line in a few cases (e.g. \citealt{tom09,pla13})   %Gierlinski et al., 1997 MNRAS, 288, 958) 
and the absence of obvious thermal components (e.g. \citealt{pou97,dov97,rem06,don07,dun10}), all potential signatures of small and remote reflecting area. %}Poutanen et al. 1997 MNRAS ,292, L2; Dove et al. 1997 ApJ ,487, 759; RM06). 

The variable disk inner radius during state transitions then becomes a natural key ingredient in most theoretical models, controlling or resulting from the spectral and timing evolution of microquasars during the outburst (e.g. \citealt{esi97, mey00a, bel05,rem06, fer06a, pet08}).\\
% . As non-exhaustif examples: it localizes the maximum of the cooling photons flux coming from the disk and entering
%the inner X-ray corona, a process that could be at the origin of the spectral variability; its evolution follows the changes of the disk magnetic radial distribution with the accretion rate, a phenomenon that could potentially explain the hysteresis cycle (see \citealt{pet08} %Petrucci et al. 2008 MNRAS, 385, L88 
%for a recent development); its associated keplerian frequency (or a fraction/proportion of it) would be associated to the frequency QPOs (either low or high e.g. \citealt{bel05,rem06}). %McClintock & Remillard 2005)
% In consequence, the QPO frequency shifts, would be linked to the variation of the disk inner radius. 

\begin{figure}
\begin{tabular}{c}
\includegraphics[width=\columnwidth]{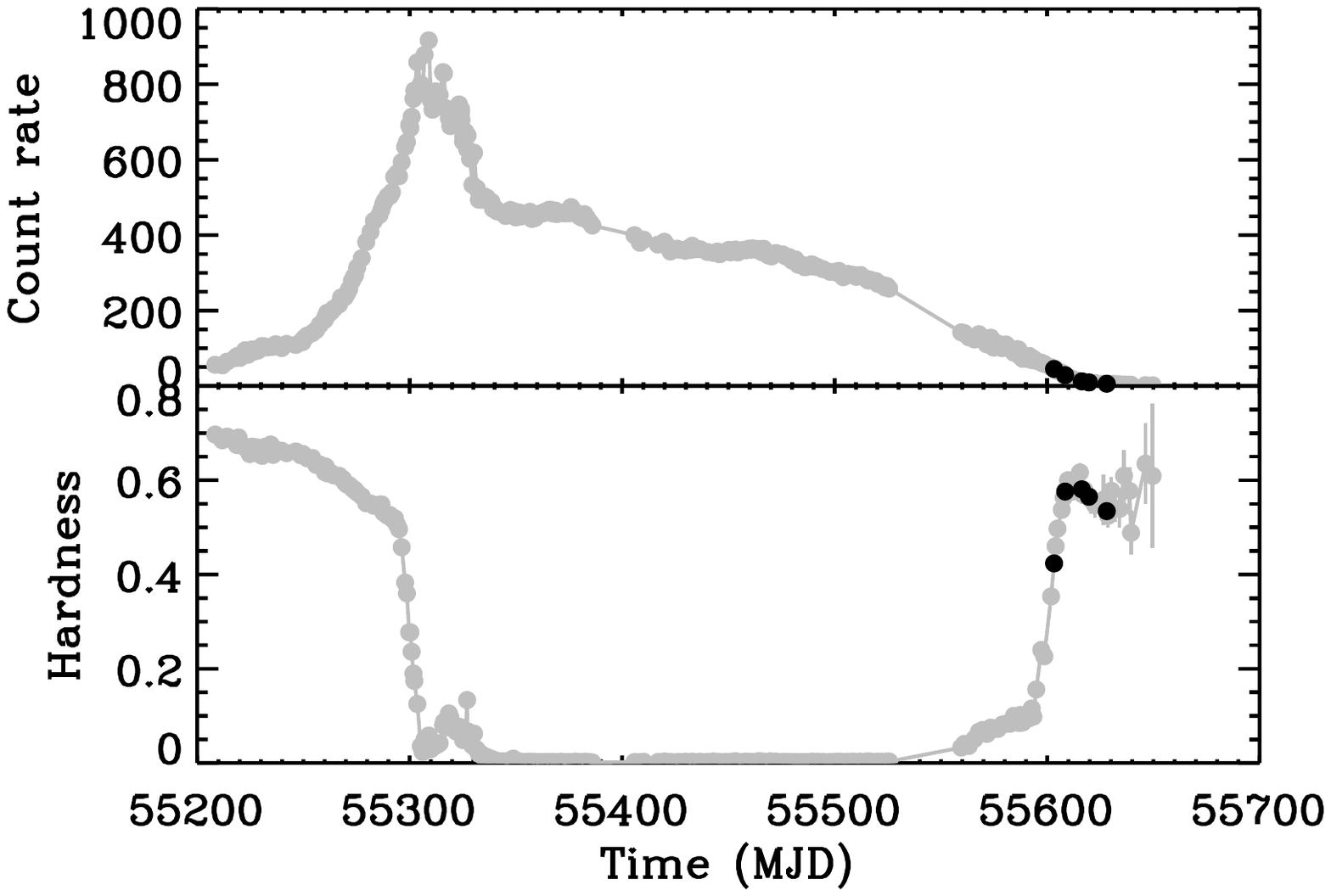}\\
\includegraphics[width=\columnwidth]{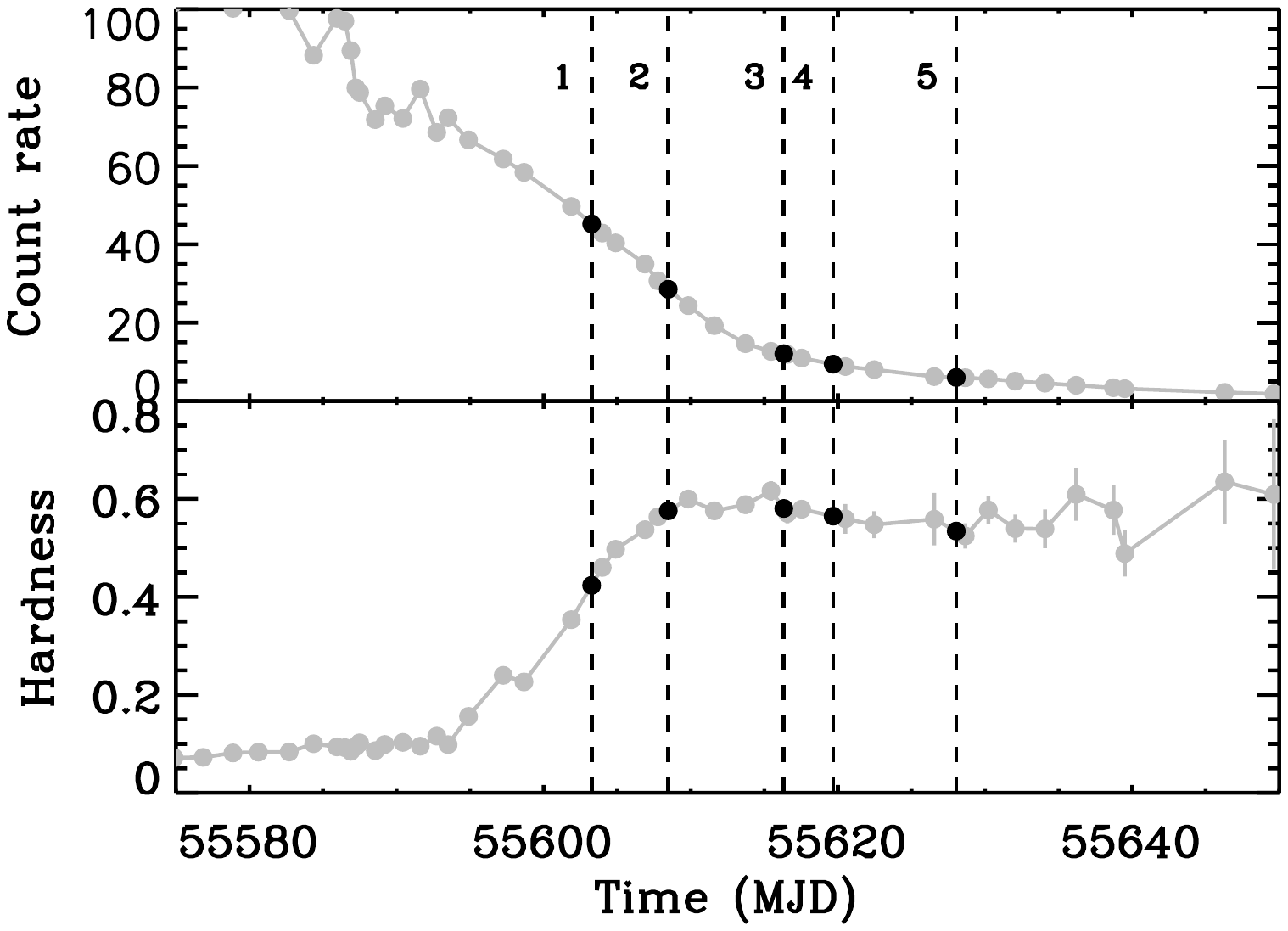}
\end{tabular}
 \caption{{\bf Top:} RXTE/PCA 3-20 keV count rate and hardness ratio light curves of the complete 2010-2011 outburst of GX 339-4. {\bf Bottom:} A zoom of the end of the outburst, with the date of the 5 Suzaku observations indicated by the black dots and the vertical dotted lines.}
\label{lc-hard}
\end{figure}

Recent XMM observations, much more sensitive, especially in the soft X-rays (i.e. below 2 keV), compared to the other existing X-ray missions, apparently ruled out the presence of a recessed disk in the hard states (e.g. \citealt{mil06a,mil06b}). %(Miller et al. 2006 ApJ, 652, L113; 2006 ApJ, 653, 525, M06 hereafter). 
The long XMM observation monitored during the 2004 outburst of GX339-4 did not agree with a simple power-law although the system was in a typical, although bright ($\sim$ 10\% L$_{Edd}$), hard state. According to the authors, a very strong soft excess as well as a broad emission feature around 6.4 keV were also present and well fitted by multicolor disk and a relativistically broadened emission line respectively. This suggests the presence of an accretion disk extending towards the vicinity of the black hole (but see below). SWIFT/XRT observations of XTE J1817- 330, during its decline to the hard state, led \cite{ryk07} to the same conclusions i.e. no disk recession. {From their SWIFT survey of stellar mass black holes, \cite{rey13} do not find evidence for large-scale truncation of the accretion disk in the hard state either, at least for X-ray luminosities larger than 10$^{-3}$ L$_{Edd}$.}\\

The estimates of inner disk radii based on continuum spectroscopy are subject however to considerable uncertainties (e.g. \citealt{mer00,zim05,cab09} hereafter C09) %Merloni et al. 2000, MNRAS, 313, 193; Zimmerman et al. 2005, ApJ 618, 832; Cabanac et al. 2009, MNRAS, 396, 1415, C09 hereafter) 
and different data analysis may give different conclusions. A recent re-analysis of the data used in \cite{mil06b} suggests that the observed broad iron line  may be an artifact due to an improper correction of the pile-up in the MOS data \citep{don10}. %(Done et al. 2009, arxiv/0911.3243). 
Suzaku pointing on the same source in its hard state has also shown that the spectrum could be consistent with a truncated disk \citep{tom09}. %(Tomsick et al. 2009, arxiv/0911.2240v1). 
More specifically, by re-analyzing the whole SWIFT/XRT dataset of XTE J1817-330, %, but allowing the column density free to vary (e.g. \citealt{ooe97} but see \citealt{mil09}) %Ooesterbroek et al., 1997, see however Miller et al. 2009), 
C09 does observe a slight increase of the disk radius. %  and note that this is also the case when NH is fixed). 
This occurs apparently when the 2 -10 keV luminosity decreases below $\sim$ 5 $\times 10^{-3}L_{Edd}$\footnote{Assuming a 10 solar masses black hole.}. These authors analyzed other sources in the same way and obtain similar, though less significant, results.

The reality of the disk recession is clearly an information of prime importance that we crucially need if we want to understand the physics of the state transitions. Confirming, and precisely measuring, this recession (if any) should allow to constrain and refine most of the present theoretical models. On the other hand, the absence of recession will imply to strongly reconsider our present understanding of the microquasar phenomenon.\\ 

We present in this paper a Suzaku campaign on the microquasar GX 339-4 aiming at catching the recession, if any, of the accretion disk during a soft-to-hard state transition. Sect. \ref{obs} detailed the observation and data treatment and Sect. \ref{datana} the data analysis. While the constraints on the disk inner radius, discussed in Sect. \ref{secrec}, prevent any clear conclusions concerning its recession, the observation of ionized absorbing features in the soft X-rays may suggest a disk wind whose properties may evolve during the transitions. Theses results are discussed in Sect. \ref{diskwind} before concluding.

\section{Observations and data treatment}
\label{obs}
\subsection{Suzaku Observations}
\gx was observed five times ($\sim$20 ks each) by Suzaku at the end of its last outburst in February 2011, as soon as the source became visible by the satellite. The 5 observations were separated by a few days in order to follow the spectral evolution of the object all along its transition back to the hard state. The log of these observations is detailed in Tab. \ref{tab1}, with the corresponding dates. The RXTE/PCA 3-20 keV light curve of the complete 2010-2011 outburst of GX 339-4 is plotted at the top of Fig. \ref{lc-hard} together with the hardness ratio\footnote{The hardness ratio is defined as the ratio of the (5.7-9.5 keV) count rate over the (2.9-5.7 keV) count rate}. A zoom of the last part of the outburst, with the dates of the 5 Suzaku observations indicated by the vertical dotted lines, is shown at the bottom of  Fig. \ref{lc-hard}.\\
\begin{figure}[b]
\includegraphics[width=\columnwidth,height=8cm]{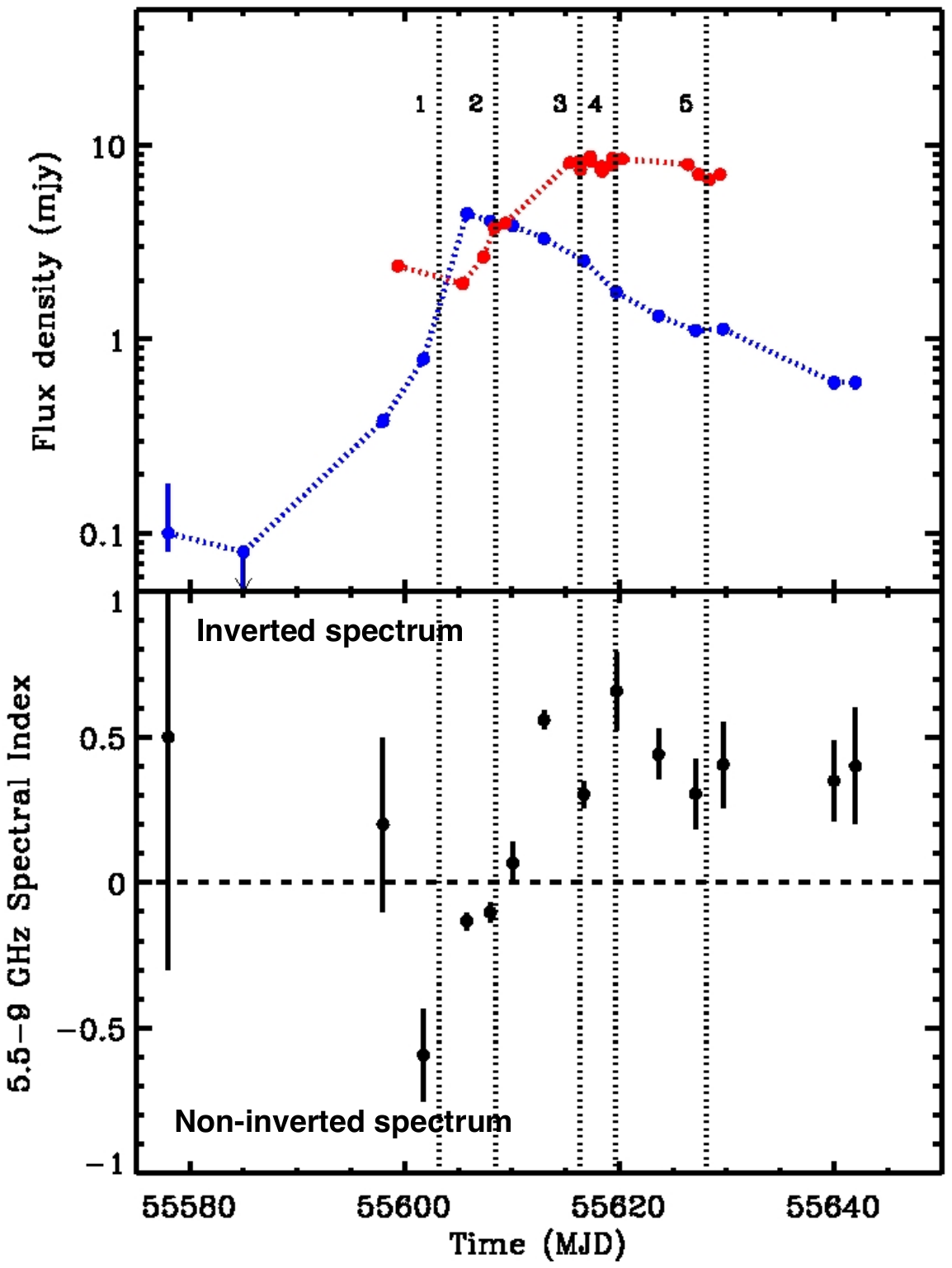}
 \caption{{\bf Top:} 5.5 GHz (blue) and H band (red) light curves. {\bf Bottom:} Light curve of the 5.5-9 GHz spectral index. {\bf A positive spectral index corresponds to an inverted spectrum generally characteristic of optically thick synchrotron emission from a stratified jet (e.g. \citealt{bla79})}. The dates of the 5 Suzaku observations are indicated by the vertical dotted lines (data from \citealt{cor13b}).} %RAJOUTER COURBE DE LUMIERE DE SUZAKU!!!}
\label{lc-radiooir}
\end{figure}

For the data treatment we use the most up-to-date calibration files and the HEASoft version 11.6.1. We run the Suzaku XIS/HXD  {\it aepipeline (V1.1.0)} tools to reprocess the data from scratch. We take care to pile-up effect in the XIS instrument by running first the {\it aeattcor2} tool which corrects Suzaku attitude data for the effects of "thermal wobbling" caused by thermal distortions of the satellite bodies \footnote{http://heasarc.gsfc.nasa.gov/ftools/caldb/help/ aeattcor2.html}. Then we run the pile-up fraction estimation tool {\it pileest} also released in the HEASoft package to estimate the amount of pileup in the XIS images and disregarded regions with $>$10\% pileup fraction during the spectral extraction. Only the first two observations suffered from pileup ({ $\sim$ 19 and 10\% for OBS1 and OBS2 respectively for the most central pixels)} and a circular region of $\sim$10" radius in the central part of the image was excluded. We added together the XIS0 and XIS3 spectra. { Unless specified in the text, the XIS spectra were all rebinned  so that
the minimum number of bins per resolution element is 5 in order to ensure a minimum number of counts per channel of 30\footnote{For the rebin procedure, we use the
tool PHARBN developed by M. Guainazzi and adapted to the XIS instrument}}. \\

To create the background files in the XIS energy range we use the ftool {\it xisnxbgen} which estimates the non X-ray background spectrum of the XIS instrument. While the Cosmic X-ray Background is expected to be low in the XIS energy band, we take it into account following the recipe indicated in the ABC guide V4.0 (p. 81-82)\footnote{http://heasarc.gsfc.nasa.gov/docs/suzaku/analysis/abc/} but renormalized to the XIS FOV corresponding to the 1/4 window mode. { The total X-ray background files (including the non-X-ray as well as the cosmic X-ray background) of the HXD/PIN %and HXD/GSO 
instruments were computed using the ftools {\it hxdpinxbpi}\footnote{We do not include the GSO data in the spectral analysis due to their very poor statistics.}. For the non X-ray background, we use the "tuned" background files distributed by the HXD team and corresponding to our observations. These "tuned" background suffers from systematic uncertainties of about 1.3\%\footnote{see http://heasarc.gsfc.nasa.gov/docs/suzaku/analysis/ watchout.html, item 16.} that have been added to the PIN data during our fitting procedure.}\\

% and {\it hxdgsoxbpi} respectively.\\
Following the {\it ABC guide V.4.0 (chapter 5.7.2)}, the normalization of the HXD/PIN %and HXD/GSO 
with respect to XIS is fixed to 1.16. % and 0.93 respectively.
The energy ranges for the XIS and HXD/PIN % and HXD/GSO 
instruments are restricted to 0.7-10 keV and  20-70 keV respectively. In the following, we name the five observations OBS1, OBS2, OBS3, OBS4 and OBS5 (see Tab. \ref{tab1}).%and 70-200 keV respectively.

\subsection{Radio and IR observations}
Simultaneous or quasi simultaneous radio observations were taken with the Australia Telescope Compact Array (ATCA) at 5.5 and 9 GHz (\citealt{cor13b}, hereafter C13). %(Corbel et al. 2012).
 %The flux in each band as well as the radio 5.5-9 GHz photon index are reported in Tab. \ref{tabflux}\\
Simultaneous or quasi simultaneous optical and IR observations were taken with  the SMARTS telescope \citep{bux12,din12}. %(Buxton et al. 2012, Dincer et al. 2012).
The corresponding light curves are reported at the top of Fig. \ref{lc-radiooir} and the radio spectral index (between the 5.5 and 9 GHz band) is plotted at the bottom of the figure.

\section{Results}
\label{datana}

\subsection{Light curves}
\gx was at the end of its outburst and, as expected, its X-ray flux decreases from OBS1 to OBS5. The count rates in the 0.7-2, 2-10, 20-70 and 70-200 keV ranges are indicated in Tab. \ref{tab1} and indeed they all decrease during the campaign. Interestingly the decrease is much more pronounced in the softer energy range than in the harder ones. The 0.7-2 keV range count rate decreases by more than a factor 10 while it decreases by less than a factor 2 in the 70-200 keV range. This clearly indicates a spectral change of the X-ray emission.\\

As shown in Fig. \ref{lc-radiooir}, the radio emission begins to increase between MJD 55590-55600 and reaches a maximum between our two first Suzaku observations ($\sim$ MJD 55605). The radio emission being generally associated with the presence of a jet, the increase of radio flux is interpreted as the reappearance of the jet while GX 339-4 is turning back to its hard state (see e.g. C13, \citealt{kal13}). This is well supported by the simultaneous increase of the radio spectral index (bottom plot on Fig. \ref{lc-radiooir}), which becomes positive after OBS2. Such inverted spectrum is characteristic of optically thick synchrotron emission from a stratified jet (e.g. \citealt{bla79}).

It is interesting to note that the RXTE/PCA hardness ratio (cf. Fig. \ref{lc-hard}) begins to increase a few days before the radio peak, the two events being clearly separated in time (see also \citealt{kal13} for other outbursts). Concerning the H band emission, it reaches a local maximum about 10 days after the radio peak, in between OBS3 and OBS4 (see the discussion in C13). 
\begin{figure}
\includegraphics[width=0.95\columnwidth]{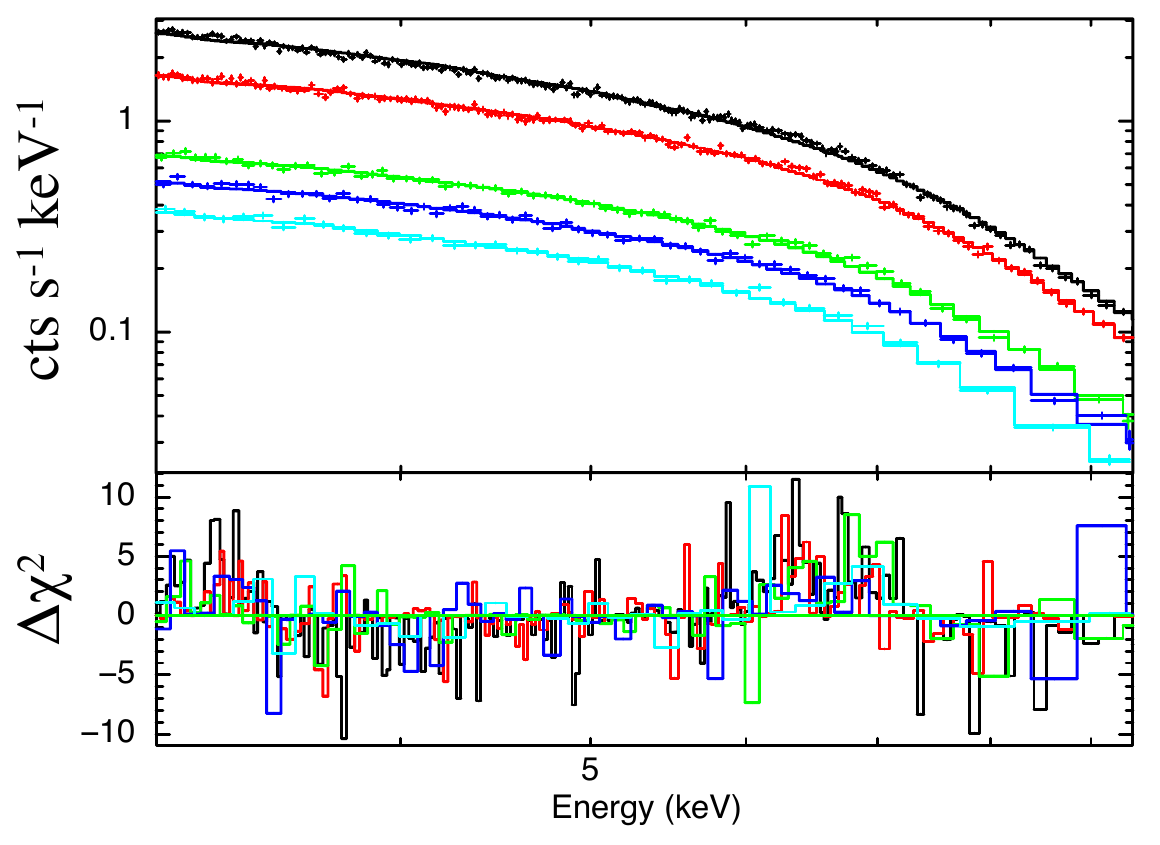}
 \caption{{\bf Top panel}: Folded spectra of the 5 Suzaku observations (OBS1 to OBS5 from top to bottom) fitted by a power law in the 3-10 keV energy range. {\bf For clarity, only the XIS0+XIS3 spectra are plotted. The have been rebinned in order to have 30$\sigma$ per bin}. The {\bf bottom panel} shows the contributions to the $\chi^2$.}
\label{nonunfolded}
\end{figure}

\begin{figure*}
\begin{tabular}{cc}
\includegraphics[width=\columnwidth,angle=0]{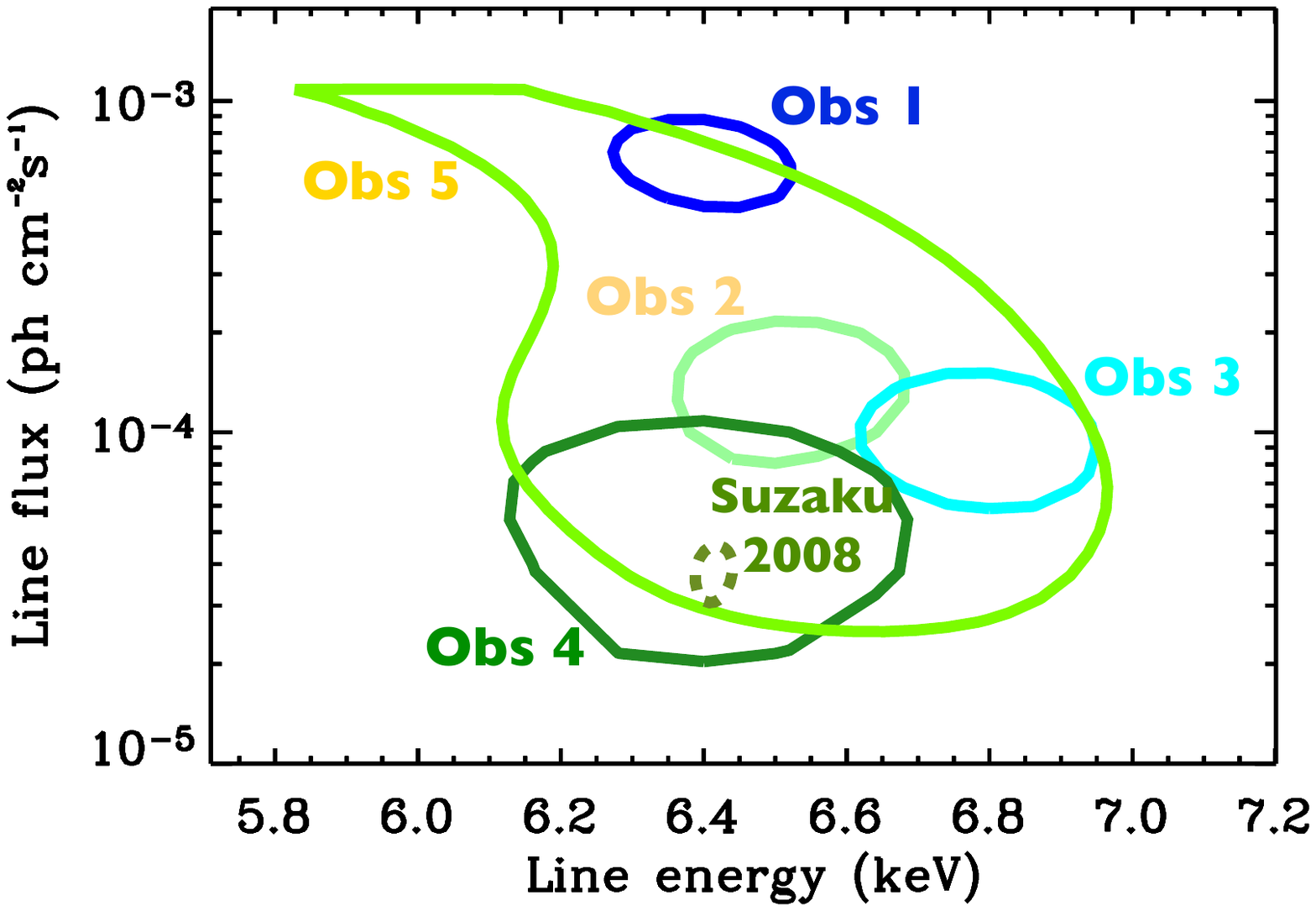}&
\includegraphics[width=\columnwidth,angle=0]{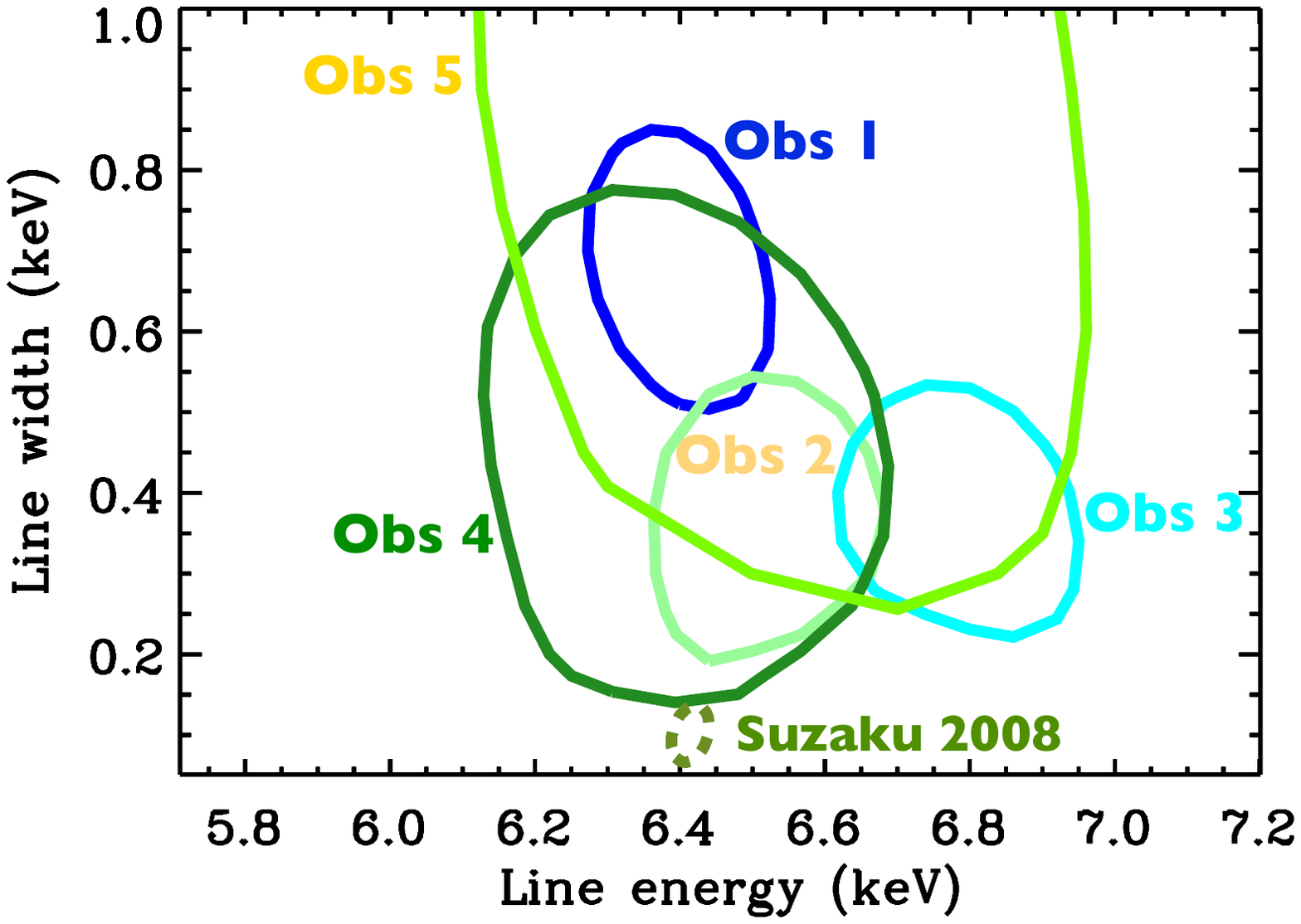}\\
\end{tabular}
 \caption{ The 90\% contour plots line flux vs. line energy ({\bf Left}) and line width vs. line energy ({\bf Right}) from the best fits of the five XIS spectra with a {\sc power law} + {\sc gaussian} model in the 3-10 keV energy range. The dashed contour corresponds to the Suzaku observation of 2008 \citep{tom09}.}
\label{contgauss}
\end{figure*}

\subsection{Spectral analysis}
\subsubsection{Above 3 keV: XIS data alone}
\label{secge3kev}

We begin our spectral analysis on the Suzaku data above 3 keV, a spectral domain known to be dominated by the primary continuum. We first fit the XIS data only between 3 and 10 keV with a simple power law model. The best fit values for the photon index are reported in Tab. \ref{tabparam} and the folded spectra are plotted in Fig. \ref{nonunfolded}. The photon index decreases rapidly between OBS1 and OBS2, from $\Gamma=$1.72$\pm$0.01 to $\Gamma=$1.56$\pm$0.01 and then stays roughly constant around 1.55 between OBS2 and OBS5. \\

Excesses close to 6-7 keV are visible especially  in OBS1, 2 and 3, suggesting the presence of an iron K$\alpha$ line. We add a gaussian component and fit the spectra again. The fit improves very significantly for OBS1 ($\Delta\chi^2$=143), less for OBS2 and 3 ($\Delta\chi^2$=29 and 30 respectively) and even less for OBS4 and 5 ($\Delta\chi^2$=14) but the improvement is always significant (at more than 99\% following the F-test\footnote{See however \cite{pro02} about the usage of the F-test in line-like features.}). The best fit parameter values are reported in Tab. \ref{tabparam}. The contour plots of the line flux vs. line energy and line width vs. line energy for the 5 observations are plotted in Fig. \ref{contgauss}.\\

The line flux is clearly decreasing between OBS1 and OBS2 (at more than 3$\sigma$). Then, between OBS2 and OBS5, it is consistent with a constant at a confidence level of $\sim$ 70\%. Concerning the line energy, its highest value is reached in OBS3 with $E_{gauss}=6.77\pm0.13$. The others measurement are consistent with a neutral iron line peaking at 6.4 keV. The significativity of the variability of the line energy during the campaign is however less than 40\%. The line width is consistent with a constant at a confidence level of $\sim$ 60\%. It has an average value of $\sim$ 440$\pm$70 eV potentially indicating some broadening.\\

\begin{figure}
\includegraphics[width=0.95\columnwidth]{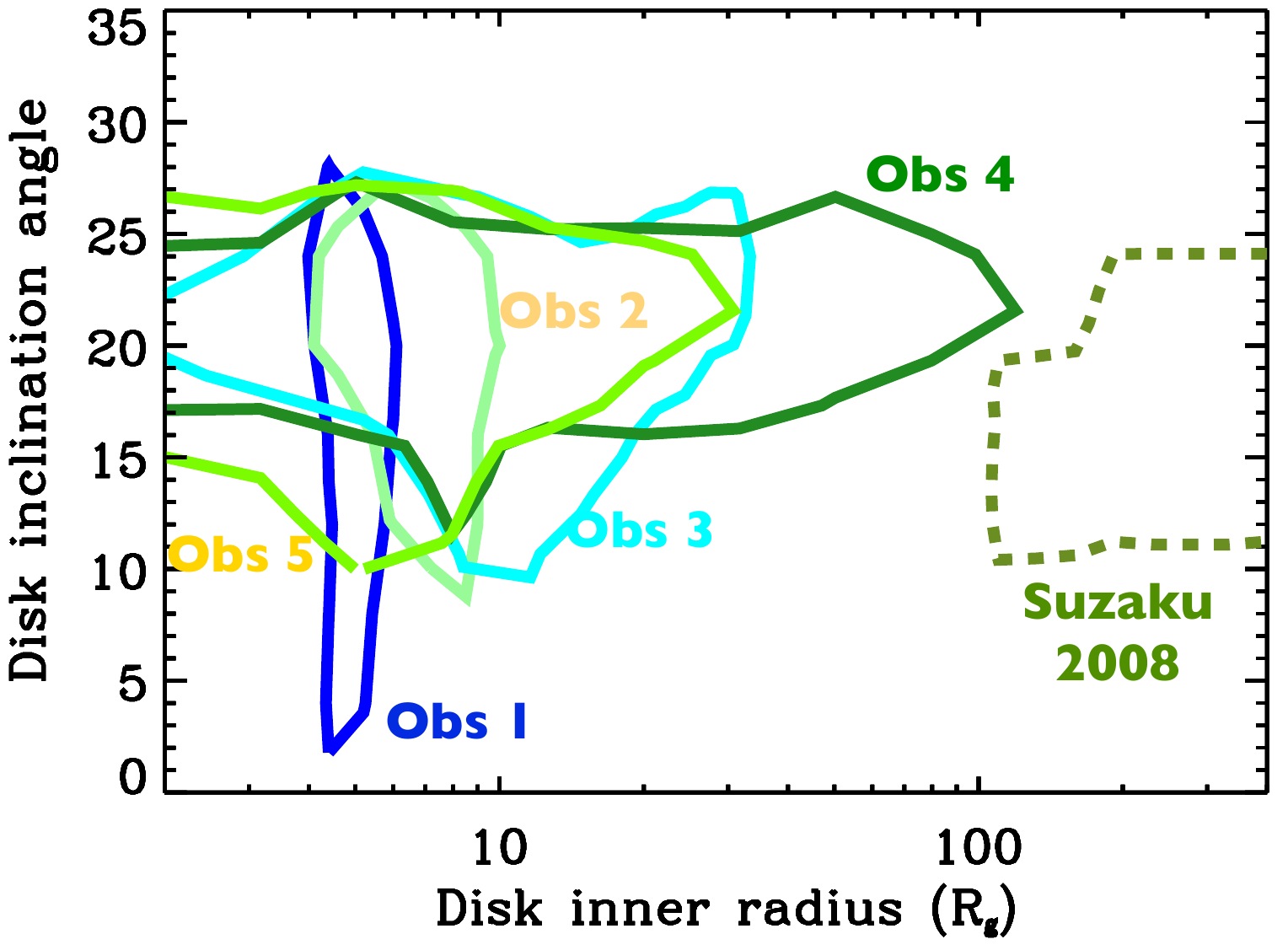}
 \caption{ The 90\% contour plots of the disk inclination angle vs. disk inner radius  obtained by fitting simultaneously the five XIS spectra  with a power law + Laor model above 3 keV. We impose the same inclination angle between the 5 models. The dashed contour corresponds to the Suzaku observation of 2008 \citep{tom09}}
\label{contlaor}
\end{figure}

To test if the line broadening could be due to relativistic effects, we replace the gaussian profile by a Laor profile which is expected if the line emission is produced by an accretion disk around a  black hole. To limit the number of free parameters, we fix the outer radius of the disk $R_{out}$ to 400 $R_g$ and the index of the radial power law dependence of the disk emissivity to 3. Then we fit the five XIS spectra simultaneously, imposing only the same disk inclination angle (which was let free to vary) for the 5 observations. All the other parameters (Laor model inner radius and normalization, power law photon index and normalization) were independent from one observation to the other and free to vary. The fit gives a best fit value of the inclination angle $i=21\pm 7$ deg, consistent with past measurements (e.g. M04, \citealt{rei08}). %Miller et al. 2004, Reig et al. 2008
The best fit values of the other parameters are reported in Tab. \ref{tabparam} and the contour plots of the disk inclination vs. the disk inner radius are reported in Fig. \ref{contlaor}.\\

The fits are statistically as good as with a gaussian profile, the line flux and equivalent width behaviors being completely consistent between the two profiles. Compared to the gaussian fits, the line energy is now consistent with a constant between the 5 observations and it is significantly higher with a mean value $\langle E_{Laor}\rangle = 6.96\pm 0.07$ keV. Concerning the disk inner radius, it is well constrained to a few Schwarzschild radii for OBS1 and OBS2 while we have only upper limits for the three last observations. While it strongly suggests the presence of the accretion disk very close to the black hole at the beginning of the campaign, the data prevent any clear conclusion concerning the disk recession.\\

% The disk inner radius  light curve is also consistent with a constant with a mean value $\langle R_{in}\rangle = 5.1\pm 0.5\ R_g$.
%

\begin{table*}
\begin{center}
\begin{tabular}{cccccccc}
\hline
\multicolumn{7}{c}{{\it power law}}\\
Obs. & $\Gamma$ & $F_{3-10\ keV}$& $L_{3-10\ keV}/L_{Edd}$& & &$ \chi^2$/dof \\
 & & ($\times$10$^{-10}$erg cm$^{-2}$ s$^{-1}$) & (\%) & &  & \\
\hline
\vspace*{-2mm}\\
1 &      1.72$\pm$0.01 & 3.35 & 0.188 & & & 429/254\\
2 &      1.56$\pm$0.01 & 2.05 & 0.115 & & & 303/254\\
3 &      1.53$\pm$0.02 & 0.89 & 0.050 & & & 297/251\\
4 &      1.53$\pm$0.02 & 0.66 & 0.037 & & & 314/245\\
5 &      1.57$\pm$0.02 & 0.47 & 0.026 & & & 220/219\\
\vspace*{-2mm}\\
\hline
\vspace*{-2mm}\\
2008&      1.52$\pm$0.01 & 0.44 & 0.024 & & & 435/254\\
\vspace*{-2mm}\\
\hline
\hline
\multicolumn{7}{c}{{\it power law+gauss}}\\
Obs. & $\Gamma$ &$E_{gauss}$ & $\sigma_{gauss}$ & $F_{gauss}$  & $EW$  & $\chi^2$/dof\\
 & &(keV) &(keV)&($\times$10$^{-4}$ph cm$^{-2}$ s$^{-1}$) & (eV) & \\
\hline
\vspace*{-2mm}\\
1 & 1.81$\pm$0.02 & 6.40$\pm$0.10 & 0.66$\pm$0.13 & 6.5$_{-1.4}^{+1.7}$ & 150$_{-35}^{+40}$& 285/251\\
2 & 1.60$\pm$0.02 & 6.49$\pm$0.12 & 0.35$\pm$0.13 & 1.5$_{-0.5}^{+0.6}$ & 55$_{-20}^{+25}$& 267/251\\
3 & 1.57$\pm$0.02 & 6.77$\pm$0.13 & 0.36$\pm$0.11 & 1.0$_{-0.4}^{+0.4}$ & 90$_{-35}^{+40}$& 263/248\\
4 & 1.56$\pm$0.03 & 6.42$\pm$0.20 & 0.36$\pm$0.22 & 0.5$_{-0.3}^{+0.4}$ & 60$_{-35}^{+35}$& 300/242\\
5 & 1.62$\pm$0.06 & 6.59$\pm$0.30 & 0.60$\pm$0.51 & 0.7$_{-0.4}^{+1.4}$ & 130$_{-105}^{+75}$& 207/216\\
\vspace*{-2mm}\\
\hline
\vspace*{-2mm}\\
2008& 1.54$\pm$0.01 & 6.41$\pm$0.02 & 0.10$\pm$0.03 & 0.4$_{-0.1}^{+0.1}$ & 65$_{-10}^{+10}$& 248/251\\
\vspace*{-2mm}\\
\hline
\hline
\multicolumn{7}{c}{{\it power law+laor}}\\
Obs. & $\Gamma$ &$E_{laor}$ & $R_{in}$ & $F_{laor}$  & $EW$  & $\chi^2$/dof\\
 & &(keV) &($R_g$)&($\times$10$^{-4}$ph cm$^{-2}$ s$^{-1}$) & (eV) & \\
\hline
\vspace*{-2mm}\\
1 & 1.80$_{-0.01}^{+0.02}$ & 7.00$_{-0.08}^{+0.13}$ & 5.0$_{-0.6}^{+0.5}$ & 6.0$_{-0.9}^{+1.1}$ & 160$_{-30}^{+40}$ & 286/250\\
2 & 1.60$_{-0.02}^{+0.02}$ & 6.84$_{-0.14}^{+0.12}$ & 6.3$_{-1.3}^{+2.7}$ & 1.9$_{-0.3}^{+0.6}$ & 80$_{-30}^{+40}$ & 266/250\\
3 & 1.58$_{-0.03}^{+0.03}$ & 7.06$_{-0.13}^{+0.16}$ & $<$35 & 1.3$_{-0.7}^{+0.5}$ & 140$_{-60}^{+60}$ & 260/247\\
4 & 1.56$_{-0.03}^{+0.03}$ & 6.82$_{-0.48}^{+0.22}$ & $<$180  & 0.7$_{-0.4}^{+0.4}$ & 90$_{-70}^{+60}$ & 300/241\\
5 & 1.61$_{-0.02}^{+0.06}$ & 7.03$_{-0.19}^{+0.38}$ & $<$25 & 0.7$_{-0.2}^{+0.6}$ & 120$_{-90}^{+130}$ & 207/215\\
\vspace*{-2mm}\\
\hline
\vspace*{-2mm}\\
2008& 1.54$_{-0.01}^{+0.01}$ & 6.43$_{-0.02}^{+0.02}$ & $>$100 & 0.38$_{-0.04}^{+0.05}$ & 65$_{-5}^{+5}$ & 248/251\\
\vspace*{-2mm}\\
\hline
\end{tabular}
\caption{\label{tabparam} Best fit of the XIS data between 3 and 10 keV with a simple power law, a power-law + gaussian line and a power-law + Laor profile emission line. In the last case, the 5 XIS spectra have been fitted simultaneously, letting all the parameters free to vary but imposing the Laor profile inclination angle to be the same between the 5 models. The best fit value for the inclination angle is i=21$_{-7}^{+5}$ deg. The errors on the inner disk radius $R_{in}$ of the Laor profile correspond to a confidence level of 90\% for 2 parameters. We have reported also the best fit parameter values for the 2008 Suzaku observation, again assuming the same inclination angle than the 5 other Suzaku observations.  With the addition of these data, the best fit value for the inclination angle then becomes $i=20_{-6}^{+3}$ deg. To compute the 3-10 keV flux in Eddington unit, we adopt an Eddington luminosity of $L_{Edd}\simeq1.3\times 10^{39}$ erg s$^{-1}$ { (i.e. we assume a 10 solar masses black hole)} and a distance to GX 339-4 of 8 kpc.}
\end{center}
\end{table*}

\begin{table*}
\begin{center}
\begin{tabular}{ccccccccccc}
%\hline
%\multicolumn{7}{c}{{\it tbnew + power law + gau}}\\
%Obs. & $\Gamma$ & & & & &$ \chi^2$/dof\\
%\hline
%1 &      1.72$\pm$0.01 & & & & & 429/254\\
%2 &      1.56$\pm$0.01 & & & & & 303/254\\
%3 &      1.53$\pm$0.02 & & & & & 297/251\\
%4 &      1.53$\pm$0.02 & & & & & 314/245\\
%5 &      1.57$\pm$0.02 & & & & & 220/219\\
\hline
%\multicolumn{11}{c}{{\it tbnew*(diskbb+power law+gau)}}\\
%Obs. & $N_h$ & $T_{in}$ & $N_{diskbb}$ & $\Gamma$ &$E_{gauss}$ & $\sigma_{gauss}$ & $F_{gauss}$  & $EW$  & $\chi^2$/dof & $\Delta\chi^2$\\
% & $\times 10^{22}$ & (keV) &   & & (keV) &(keV)& $\times 10^{-5}$ & (eV) & \\
%\hline
%\vspace*{-2mm}\\
%1 &0.46$_{-0.01}^{+0.02}$ & 0.36$_{-0.02}^{+0.02}$&647$_{-123}^{+167}$ & 1.84$_{-0.02}^{+0.03}$ & 6.44$_{-0.10}^{+0.10}$ & 0.69$_{-0.14}^{+0.18}$ & 7.0$_{-1.6}^{+2.3}$ & 165$_{-40}^{+40}$& 579/384 & 870\\
%
%2 &0.66$_{-0.04}^{+0.04}$ & 0.20$_{-0.01}^{+0.01}$& 6.6$_{-2.5}^{+3.5}\times 10^3$ & 1.68$_{-0.01}^{+0.02}$ & 6.56$_{-0.12}^{+0.12}$ & 0.38$_{-0.11}^{+0.17}$ & 1.7$_{-0.5}^{+0.7}$ & 70$_{-25}^{+20}$& 438/384 & 168\\
%
%3 &0.57$_{-0.03}^{+0.09}$ & $<$0.19&$>$300 & 1.60$_{-0.02}^{+0.03}$ & 6.79$_{-0.13}^{+0.13}$ & 0.35$_{-0.10}^{+0.12}$ & 1.0$_{-0.3}^{+0.4}$ & 90$_{-30}^{+30}$& 386/381 & 6\\
%
%4 &0.70$_{-0.07}^{+0.09}$ & 0.14$_{-0.04}^{+0.02}$&5.1$_{-3.8}^{+21.8}\times 10^3$ & 1.64$_{-0.03}^{+0.03}$ & 6.48$_{-0.20}^{+0.22}$ & 0.40$_{-0.20}^{+0.30}$ & 0.6$_{-0.3}^{+0.4}$ & 70$_{-40}^{+50}$& 426/374 & 8\\
%
%5 &0.77$_{-0.11}^{+0.16}$ & $<0.50$ & $>$0 & 1.66$_{-0.14}^{+0.05}$ & 6.66$_{-0.30}^{+0.30}$ & 0.55$_{-0.25}^{+0.50}$ & 0.6$_{-0.3}^{+0.6}$ & 110$_{-70}^{+80}$& 355/343 & 1\\
%\vspace*{-2mm}\\
%\hline
%\vspace*{-2mm}\\
%2008&0.51$_{-0.01}^{+0.01}$ & 0.06$^{+0.02}_{-0.02}$ & $<$4.2$\times 10^8$ & 1.62$_{-0.01}^{+0.01}$ & 6.42$_{-0.02}^{+0.02}$ & 0.10$_{-0.03}^{+0.03}$ & 3.8$_{-0.6}^{+0.6}$ & 65$_{-10}^{+10}$& 407/384 & 19\\
%\vspace*{-2mm}\\
%\hline
%\end{tabular}
\multicolumn{11}{c}{{\it tbnew*(diskbb+power law+laor)}}\\
Obs. & $N_h$ & $T_{in}$ & $N_{diskbb}$ & $\Gamma$ &$E_{laor}$ & $R_{in,laor}$ & $F_{laor}$  & $EW$  & $\chi^2$/dof & $\Delta\chi^2$\\
 & $\times 10^{22}$ & (eV) & $\times 10^{3}$  & & (keV) &($R_g$)& $\times 10^{-4}$ & (eV) & \\
\hline
\vspace*{-2mm}\\
1 &0.47$_{-0.01}^{+0.01}$ & 330$_{-10}^{+10}$&1.0$_{-0.2}^{+0.2}$ & 1.84$_{-0.01}^{+0.02}$ & 7.04$_{-0.07}^{+0.06}$ & 5.0$_{-0.5}^{+0.9}$ & 5.9$_{-0.9}^{+1.0}$ & 160$_{-30}^{+30}$&  593/384 & 1332\\
2 &0.63$_{-0.03}^{+0.04}$ & 210$_{-10}^{+10}$&5.3$_{-1.9}^{+2.8}$ & 1.69$_{-0.01}^{+0.02}$ & 6.89$_{-0.09}^{+0.12}$ & 6.8$_{-1.6}^{+2.3}$ & 2.1$_{-0.5}^{+0.7}$ & 80$_{-30}^{+30}$&  441/384 & 173\\
3 &0.57$_{-0.02}^{+0.10}$ & $<$160 %0.07$_{-0.05}^{+0.09}$
&$>$10.0 %733.8$_{-622.0}^{+1e5}$ 
& 1.61$_{-0.03}^{+0.03}$ & 7.07$_{-0.24}^{+0.15}$ & 8.2$_{-3.0}^{+24.4}$ & 1.2$_{-0.5}^{+0.5}$ & 110$_{-40}^{+40}$& 386/381 & 6\\
4 &0.59$_{-0.05}^{+0.04}$ & 210$_{-10}^{+10}$&5.3$_{-1.9}^{+2.8}$ & 1.69$_{-0.01}^{+0.02}$ & 6.89$_{-0.09}^{+0.12}$ & 6.8$_{-1.6}^{+2.3}$ & 2.1$_{-0.5}^{+0.7}$ & 80$_{-30}^{+30}$&  423/374 & 7\\
5 &0.67$_{-0.04}^{+0.03}$ & $<8\times 10^3$&$<$0.3 & 1.63$_{-0.10}^{+0.20}$ & 7.17$_{-0.47}^{+0.27}$ & $<$30 & 0.7$_{-0.5}^{+0.4}$ & 130$_{-90}^{+120}$&  358/340 & 0\\

\vspace*{-2mm}\\
\hline
\vspace*{-2mm}\\
2008&0.51$_{-0.01}^{+0.01}$ & 60$^{+20}_{-20}$ & $<$3.9$\times 10^8$ & 1.62$_{-0.01}^{+0.01}$ & 6.44$_{-0.02}^{+0.02}$ & $>$210 & 0.4$_{-0.1}^{+0.1}$ & 70$_{-10}^{+10}$& 408/384 & 19\\
\vspace*{-2mm}\\
\hline
\end{tabular}
\caption{\label{tabparam2} Best fit of the XIS data between 0.7 and 10 keV with a diskbb + power-law + Laor line  and a photo-electric absorption (model {\sc tbnew}, \citealt{wil00}). The line flux is in units of ph cm$^{-2}$ s$^{-1}$. We assume a 10 solar masses black hole for $R_g$. We report also the $\Delta\chi^2$ fit improvement due to the addition of the multicolor disk component.}.
\end{center}
\end{table*}

%%%%%%%%%%%%%%
\subsubsection{0.7-10 keV: XIS data alone}
\begin{figure}
\includegraphics[width=\columnwidth]{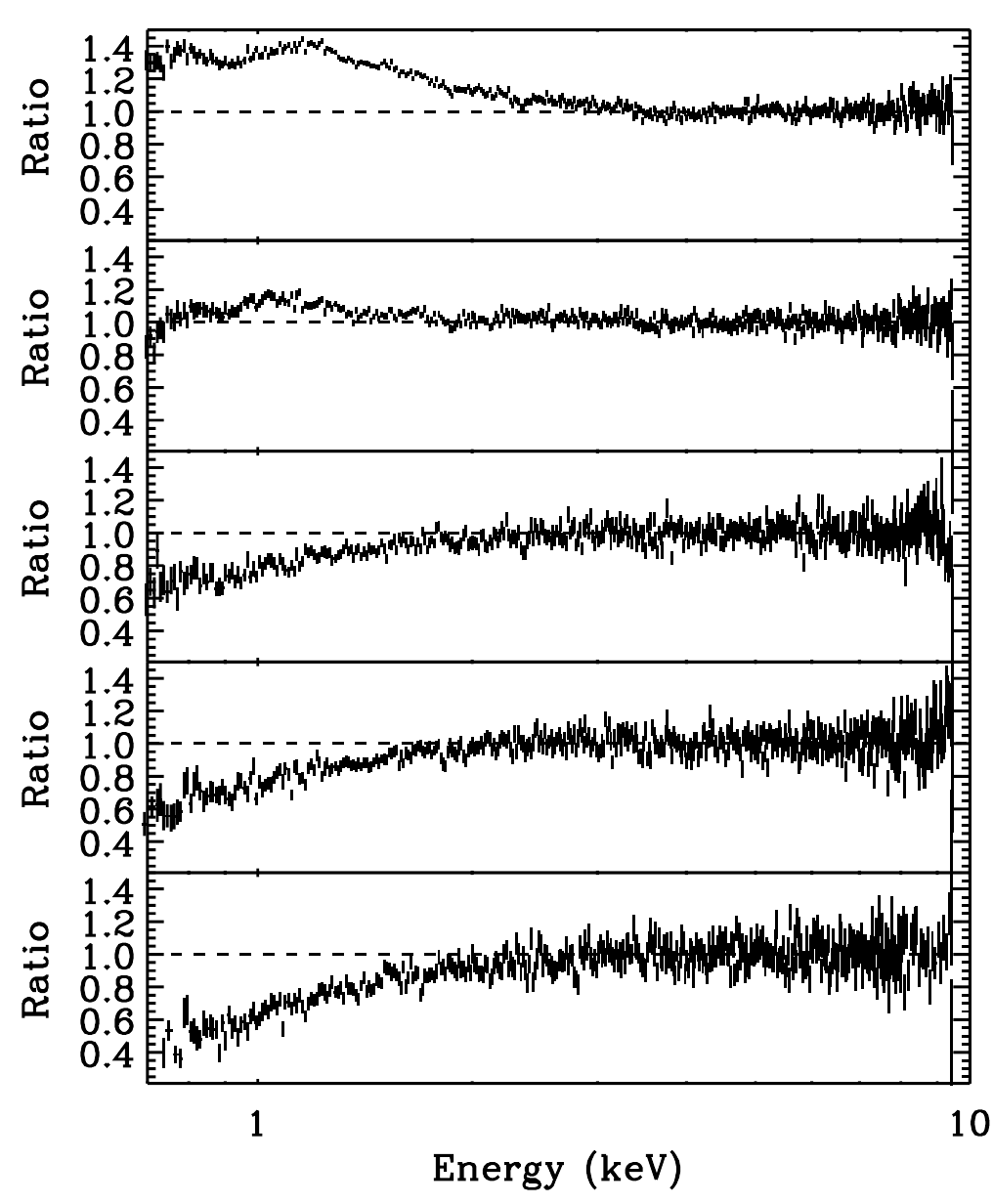}
 \caption{Ratios data/model of the five { XIS0+XIS3} spectra for OBS1 (top) to OBS5 (bottom). The model is a {\sc power law + laor} fitted above 3 keV and then extrapolated down to 0.7 keV. The Hydrogen column density is fixed to 6$\times 10^{21}$cm$^2$. We use {\sc tbnew} to model the X-ray absorption \citep{wil00}.}
\label{softxextrapol}
\end{figure}

%%%%%%%%%%%%%%
\label{secxisge07}
We now include the 0.7-3 keV data in the fitting procedure. The ratios data/model of the five XIS spectra obtained after extrapolation of the best fit {\sc power law + laor} model down to 0.7 keV are plotted in Fig. \ref{softxextrapol}. A photo-electric absorption (model {\sc tbnew}\footnote{All modeling was performed using the "wilm" abundances (Wilms, Allen \& McCray 2000) with the vern cross-sections (Verner et al. 1996)} in {\sc xspec}) was added but the Hydrogen column density was fixed to 6$\times 10^{21}$cm$^2$ i.e. the typical  galactic $N_H$ observed in the direction of GX 339-4 (e.g. \citealt{zdz04b,cad11}). %Zdziarski et al. 2004; Cadolle Bel et al. 2011)
The soft X-ray part of the spectra varies in time, in excess above the power law extrapolation in OBS1 and OBS2, but in deficit in the three last observations. Letting the column density free to vary improves the fits, but they are still statistically unacceptable especially in the case of OBS1 and OBS2. \\ 

The soft X-ray excess could be the signature of the optically thick accretion disk component which dominates the X-ray emission in the soft state, then smoothly disappearing from OBS1 to OBS5 when the source re-enters in the hard state. To test this hypothesis, we have added a multicolor disk component {\sc diskbb} in our fits. { The column density is still free to vary independently for the 5 observations}\footnote{{ Imposing $N_h$ to the same value for the 5 observations give a much worse fit with $\Delta\chi^2$=171 for 4 less dof.}}. The best fit parameters are reported in Tab. \ref{tabparam2}. The addition of a multicolor disk component improves strongly the fit of OBS1 and OBS2 with $\Delta\chi^2$=1332 and 173 respectively. This component is not statistically required however in the three last observations potentially suggesting the disappearance of the disk component during the state transition. \\

\begin{figure}
\includegraphics[width=\columnwidth]{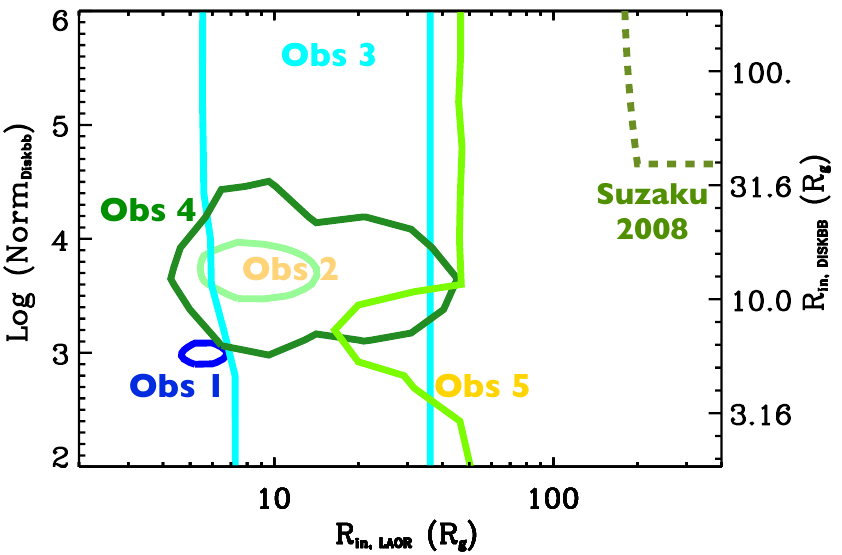}
\includegraphics[width=\columnwidth]{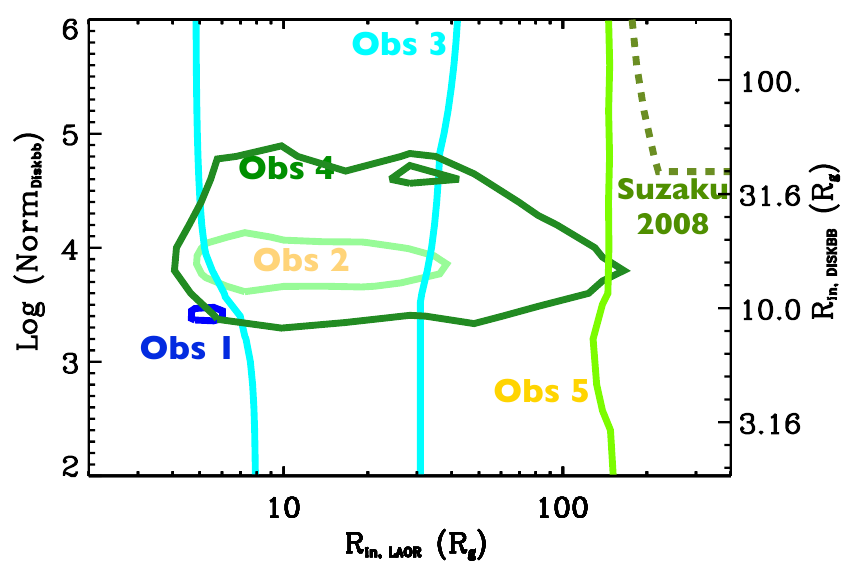}
 \caption{ Contour plots of the disk normalization (left scale) and the corresponding disk inner radius $R_{in,{{diskbb}}}$ (right scale) vs. the disk inner radius $R_{in,{{laor}}}$ deduced from the {\sc laor} profile. The five XIS spectra have been fitted with a {\sc diskbb + laor} model and a {\sc power law} (top) or a comptonization model {\sc comptt} (bottom) for the continuum. We assume a disk inclination of 20 deg., a distance of 8 kpc and a black hole mass of 10 solar masses. { For OBS5, the 90\% confidence area is on the left of the yellow line.} The dashed contour corresponds to the Suzaku observation of 2008 \citep{tom09}.}
\label{contdiskbb}
\end{figure}

We have reported on Fig. \ref{contdiskbb} the contour plots, for the 5 observations,  of the {\sc diskbb} normalization vs. $R_{in,{{laor}}}$ the best fit value of the inner disk radius of the {\sc laor} profile. Assuming an inclination angle of 21 deg., a distance of 8 kpc and a black hole mass of 10 solar masses, the {\sc diskbb} normalizations can also be converted { to an "apparent" disk inner radii $R_{in,app}$.  Taking into account a color temperature correction factor $f_{col}=1.7$ \citep{shi95,kub98,dav06,don12},  we deduce the "true" inner radius $R_{in,{{diskbb}}}={f_{col}}^2 R_{in,app}$. The values of  $R_{in,{diskbb}}$ have been reported on the right axis of Fig. \ref{contdiskbb}a}. \\

{ Both estimates of the disk inner radius (i.e. from {\sc diskbb} or {\sc laor}) roughly agree one with each other. Note a few differences however: $R_{in,{{laor}}}$ are consistent between OBS1 and OBS2 while $R_{in,{{diskbb}}}$ of OBS1 is smaller and inconsistent with the OBS2 value. In both cases however, the inner radius is found to be lower than or of the order of 10 $R_g$. In OBS3 $R_{in,{{laor}}}$  is constrained to be in between $\sim$5 and $\sim$40 $R_g$ while $R_{in,{{diskbb}}}$ is unconstrained. In OBS5 $R_{in,{{laor}}}$ is upper limited to $\sim$50 $R_g$ while $R_{in,{{diskbb}}}$ is unconstrained. }
We reach however the same conclusions for the evolution of the disk inner radius, $R_{in}$ being small in OBS1 and OBS2 ($<$ 10$R_g$), but then the contours for OBS3, OBS4 and OBS5 becomes too large to constrain its behavior. \\

%Compared to the Laor profile fits, we have however a clear increase of $R_{in}$ between OBS1 and OBS2 associated with a clear decrease of the disk inner temperature. This trend however could be model dependent. A power-law does not roll over at low energies while a realistic comptonization model should. This should have some i
{ We have checked that these results do not significantly change when using a more physical model (via e.g. up-scattering of disk photons) for the high energy continuum like {\sc comptt} in {\sc xspec}. Such models indeed differ to a power-law shape especially in the low energy portion of the spectrum where a low energy roll-over is expected. The corresponding contour plots obtained when fitting with {\sc comptt} are reported in Fig. \ref{contdiskbb}b. The main differences with Fig. \ref{contdiskbb}a are larger upper limits on $R_{in,{{laor}}}$ in OBS4 and OBS5.}\\

\begin{figure}
\includegraphics[width=\columnwidth]{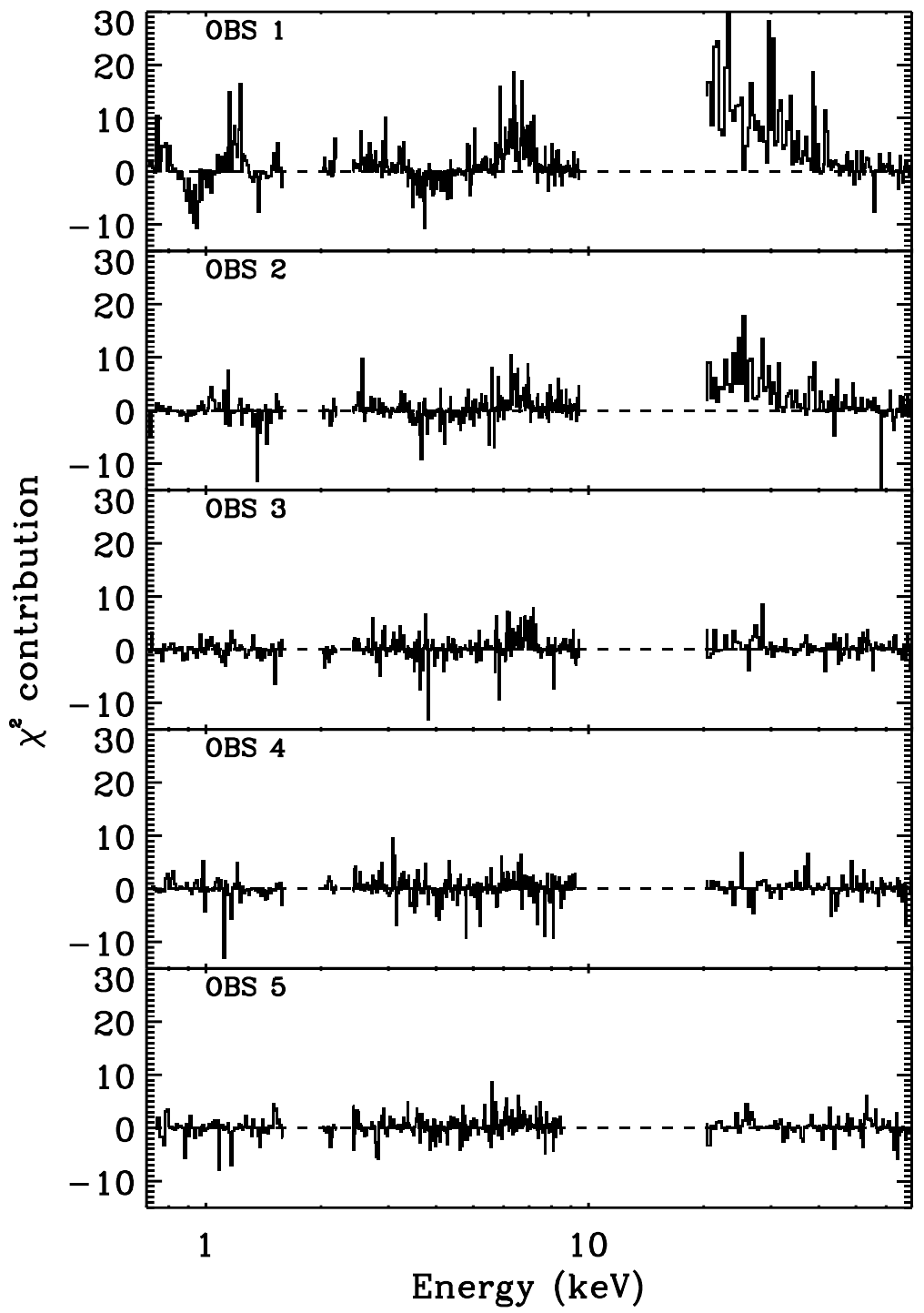}
 \caption{$\chi^2$ contribution in the 0.7-70 keV range when extending the best fit model {\sc tbnew$\times$ (diskbb+ power law + laor)} (see Tab. \ref{tabparam2}) of the XIS data above 10 keV in the PIN energy range. The {\sc laor} component has been suppressed in these plots. Strong residuals are visible near 6.4 keV and above 10 keV especially for OBS1 and OBS2.}
\label{chi2contrib}
\end{figure}

Note that the fits reported in Tab. \ref{tabparam2} are relatively bad especially for OBS1 and indeed residuals are visible in the soft part ($<$ 2 keV, see Fig. \ref{chi2contrib} and \ref{fiteqpair}) of the spectrum indicating that a multicolor disk component alone does not fit the data correctly. \\

%{\bf Note also that a more physical model for the high energy continuum, via e.g. up-scattering of disk photons, differs to a power-law shape especially in the low energy portion of the spectrum where a low energy roll-over is expected. We checked that the previous results do not significantly change when using such kind of model (like {\sc comptt} in {\sc xspec}, see Fig. \ref{contdiskbb}b)},the only effect being a larger upper limit on $R_{in,{{laor}}}$ in OBS4 and OBS5. \\

\begin{figure}
\includegraphics[width=\columnwidth]{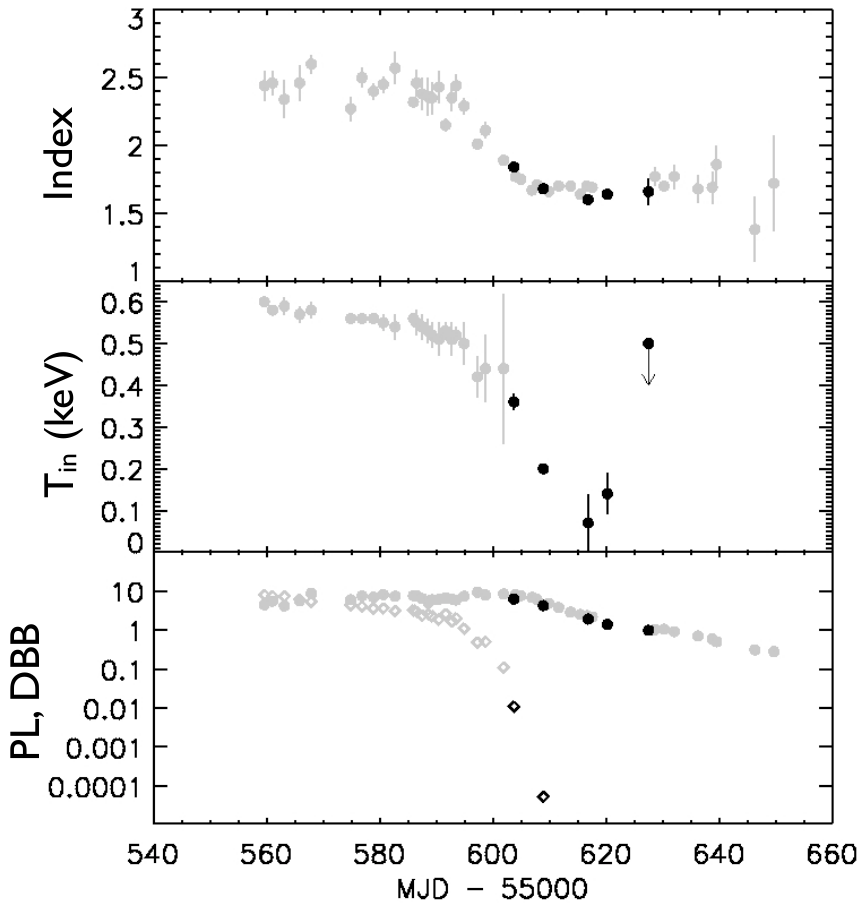}
 \caption{Evolution of the power law photon index (top), the inner disk temperature $T_{in}$ (middle) and the power-law and disk flux (bottom, filled and empty  circles respectively) in the 3-25 keV energy band in units of $10^{-10}$ erg cm$^{-2}$s$^{-1}$. The black  symbols are the results of this campaign while the gray ones have been obtained by \cite{din12}.  The model used is {\sc tbnew$\times$ (diskbb+ power law + gaussian)} in the 0.7 - 10 keV range.}
\label{compardincer}
\end{figure}

In Fig. \ref{compardincer}, we have reported our best fit results with those obtained by \cite{din12}. These authors use the RXTE/PCA data during the same outburst decay (from MJD 55559.58 to MJD 55649.64). They used also a {\sc diskbb + power law} model, as well as neutral absorption (due to their limited band pass at low energy, they fixed the hydrogen column density at 5 $\times 10^{21}$ $cm^{-2}$)\footnote{They also added a smeared edge at a fixed energy of 10 keV to fit the iron K absorption edge seen around 7.1 keV.}  so that our results can be safely compared to theirs.  As can be seen on Fig. \ref{compardincer}, both best fits parameter values agree very well one with each other. Compared to RXTE however, the lower limit of the energy band of the XIS suzaku instrument allows to follow the disappearance of the {\sc diskbb} component down to lower flux and lower inner disk temperature.

\subsubsection{Comparison with the 2008 Suzaku observation}
GX 339-4 was observed by Suzaku in 2008 during a long exposure where flux and spectral index values were very close to the one of OBS5 (\citealt{tom09}, T09 hereafter). It is worth noting however that this observation was made 1.6 years after the peak of
its 2007 outburst and that the source was in a persistent, but faint, hard state (e.g. \citealt{rus08b,kon08}). In comparison, in 2011 GX 339-4 turned back in the hard state since $\sim$ 1 month, when OBS5 was made, and the source flux was clearly in the decreasing phase of the end of the outburst.

The long exposure (the combined XIS0+XIS3 exposure time is of the order of 210 ks) results in a high statistics compared to our own observations. From the fit of the iron line profile, T09 showed that the data agreed with a truncated accretion disk with $R_{in}>30 R_g$ at 90\% confidence for a disk inclination of 0 deg, and $R_{in}>70 R_g$ for a disk inclination of 20 deg. \\ 

We have re-analyzed these data following the procedure detailed in Sect. \ref{obs}. We have fitted them in the 3-10 keV energy range with the {\sc power law + gau} model first. The corresponding contour plots of the line flux and line width versus line energy are overplotted in Fig. \ref{contgauss} and the best fit parameter values are reported in Tab. \ref{tabparam}. If the spectral shape and flux are in good agreement with OBS5, the line width is clearly inconsistent between the two observations, indicating intrinsic differences (geometry? ionisation state?) of the reflecting material.\\

Then we have fitted these data, still in the 3-10 keV energy range, but with the {\sc power law + laor} model, either separately or simultaneously to our 5 Suzaku pointings, imposing, in the latter case, the inclination angle to the same value for each data set. The best fit results are very similar in both cases and are in agreement with those of T09 (see last row of Tab. \ref{tabparam}). We find for 2008 an inner radius $>$ 100 $R_g$ and a best fit inclination angle $i=20_{-6}^{+3}$ degrees consistent with our previous value obtained without the use of these data (see Sect. \ref{secge3kev}). We have overploted the 90\% contour plot of the disk inclination angle vs. disk inner radius of the 2008 observation in dashed line in Fig. \ref{contlaor}. {\bf While only at a 2$\sigma$ level, the fact that it does not overlap with the 90\% contours obtained for our 2011 observations strongly suggests that the constraints on the disk inner radius deduced from the {\sc laor} model are unconsistent between 2008 and 2011}.\\

Following the fitting procedure of the previous sections, we have also fitted the 2008 data set in the 0.7-10 keV range adding now a {\sc diskbb} component to the model. The best fit results are reported in the last row of Tab. \ref{tabparam2} and the corresponding contour plots of the {\sc diskbb} normalization vs. inner disk radius have been overploted in dashed line in Fig. \ref{contdiskbb}. The constraints on $R_{in}$ deduced from the {\sc diskbb} normalization give now a lower limit of $\sim$30$R_g$ (to be compared with the lower limit of $\sim$ 200 $R_g$ from the $Laor$ fit of the iron line), in agreement with our (unconstrained) contour plots of OBS3 and OBS5. These constraints on $R_{in}$ from the {\sc diskbb} component have to be taken with caution however given that the disk component is poorly constrained in these observations.

\begin{table*}
\begin{center}
\hspace*{-0.5cm}
\begin{tabular}{ccccccccccccc}
\hline
\multicolumn{13}{c}{{\it tbnew*(diskbb+po+kdblr*reflionx)}}\\
Obs. & $N_h$ & $T_{diskbb}$ &  $N_{diskbb}$ & $F_{diskbb}$ &  $\Gamma$ & $F_{po}$ &  $R_{in}$ & $\xi_{ref}$ & $F_{ref}$  & $\displaystyle\frac{L_{bol}}{L_{Edd}}$ & $\chi^2$/dof\\ %& $L_{1-100\ keV}^{unabs}/L_{Edd}$ 
%\hline
 & $\times 10^{22}$ &  $eV$ &  $\times 10^4$&  $\times 10^{-11}$&  & $\times 10^{-10}$ &  & $R_{g}$ & & $\times 10^{-11}$ &  \% & \\ %& $L_{1-100\ keV}^{unabs}/L_{Edd}$ 
\hline
\vspace*{-2mm}\\
1 & 0.42$_{-0.01}^{+0.01}$ & 430$_{-20}^{+20}$&  0.03$_{-0.01}^{+0.01}$& 11.8 & 1.73$_{-0.01}^{+0.01}$ & 13.7  & 15$_{-5}^{+30}$&1000$_{-100}^{+200}$& 11.3 & 0.89& 806/517\\
2 & 0.64$_{-0.06}^{+0.03}$ & 220$_{-10}^{+20}$&  0.33$_{-0.18}^{+0.28}$& 3.5& 1.67$_{-0.02}^{+0.01}$ & 9.0  & 40$_{-25}^{+240}$&230$_{-15}^{+30}$& 11.1 & 0.56& 601/517\\
3 & 0.67$_{-0.06}^{+0.09}$ & 110$_{-40}^{+30}$&  6.10$_{-5.10}^{+216.80}$& 0.4& 1.59$_{-0.03}^{+0.02}$ & 4.0  & $>$70 &1100$_{-400}^{+800}$& 3.8 & 0.25& 522/514\\
4 & 0.75$_{-0.06}^{+0.07}$ & 125$_{-30}^{+20}$& 1.84$_{-1.31}^{+5.81}$& 0.4 & 1.63$_{-0.03}^{+0.02}$ & 3.0  & $>$10 &790$_{-600}^{+5700}$& 2.1 & 0.20& 554/507\\
5 & 0.82$_{-0.13}^{+0.25}$ & $<$240& $<2.13$ &$<$2.6 & 1.63$_{-0.18}^{+0.04}$ & 2.1  & $>$30 &1400$_{-1100}^{+3900}$& 2.0 & 0.13& 483/476 &\\
\vspace*{-2mm}\\
\hline
\vspace*{-2mm}\\
2008 &0.79$_{-0.02}^{+0.02}$ & 80$_{-20}^{+20}$ &  64.64$_{-57.09}^{+960.70}$&  0.1& 1.65$_{-0.01}^{+0.01}$ & 1.9 &  $>$180 & 20$_{-10}^{+30}$ & 3.0 & 0.13 &527/517\\
\vspace*{-2mm}\\
\hline
\end{tabular}
\hspace*{-0.5cm}
\begin{tabular}{cccccccccccccc}
\hline
\multicolumn{13}{c}{{\it tbnew*(diskbb+eqpair+kdblr*reflionx)}}\\
Obs. & $N_h$ & $T_{diskbb}$ & $F_{diskbb}$ & $l_h/l_s$ & $\tau$  & $T_{bb}$ & $F_{eqpair}$ & $R_{in}$ & $\xi_{ref}$ & $F_{ref}$  & $\displaystyle\frac{L_{bol}}{L_{Edd}}$ & $\chi^2$/dof \\ %& $L_{1-100\ keV}^{unabs}/L_{Edd}$ 
%\hline
 & $\times 10^{22}$ &  $eV$ & $\times 10^{-11}$ &  &  & $eV$ & $\times 10^{-10}$ & $R_{g}$ & & $\times 10^{-11}$ &  \% & \\ %& $L_{1-100\ keV}^{unabs}/L_{Edd}$ 

\hline
\vspace*{-2mm}\\
1 &0.46$_{-0.01}^{+0.01}$ & 310$_{-10}^{+10}$ & 16.0 & 5.1$_{-0.1}^{+0.2}$ & 0.6$_{-0.3}^{+0.3}$ & 590$^{+10}_{-30}$ & 11.9 & 7.0$_{-1.3}^{+1.1}$ & 2000$_{-200}^{+100}$ & 28.8 & 0.91 & 704/515\\
2 &0.62$_{-0.06}^{+0.07}$ & 240$_{-20}^{+10}$ & 7.0 & 5.7$_{-0.1}^{+0.3}$ & 2.3$_{-0.1}^{+0.1}$ & 470$^{+20}_{-20}$ & 9.3 & 32$_{-19}^{+33}$ & 270$_{-70}^{+185}$ & 7.8 & 0.60 & 544/515\\
3 &0.75$_{-0.03}^{+0.03}$ & 110$_{-30}^{+30}$ & 0.8 & 9.6$_{-2.1}^{+1.9}$ & 2.1$_{-1.4}^{+1.6}$ & 210$^{+50}_{-95}$ & 4.0 & $>$70 & 1180$_{-480}^{+480}$ & 4.6 & 0.25 & 519/512\\
4 &0.78$_{-0.25}^{+0.08}$ & 175$_{-40}^{+100}$ & 1.3 & 6.4$_{-0.5}^{+0.5}$ & 2.4$_{-0.2}^{+2.2}$ & 390$_{-100}^{+50}$ & 3.1 & $>$10 & 910$_{-520}^{+2100}$ & 1.7 & 0.19 & 546/505\\
5 &0.93$_{-0.14}^{+0.15}$ & 130$_{-40}^{+30}$ & 0.7 & 7.7$_{-0.6}^{+4.0}$ & 1.9$_{-1.4}^{+2.0}$ & 240$^{+60}_{-60}$ & 2.1& $>$30 & 1450$_{-1150}^{+3590}$ & 2.3 & 0.13 & 475/474\\
%3 &0.51$_{-0.03}^{+0.08}$ & $<800$ & $>$0 & 6.2$_{-2.4}^{+0.7}$ & 2.7$_{-0.3}^{+2.0}$ & & 680$^{+100}_{-60}$ & 0.5$_{-0.2}^{+0.2}$ & $>$30 & 1800$_{-480}^{+1050}$ & 0.12$_{-0.03}^{+0.03}$ & 520/512\\
%4 &0.76$_{-0.07}^{+0.09}$ & 180$^{-20}_{+40}$ & 3.6$_{-1.9}^{+4.4}$ & 6.4$_{-0.5}^{+0.3}$ & 2.5$_{-0.3}^{+1.9}$ & & 390$^{+50}_{-60}$ & 3.4$_{-0.7}^{+1.4}$ & $>$20 & 980$_{-310}^{+870}$ & 0.06$_{-0.02}^{+0.07}$ & 551/505\\
%5 &0.66$_{-0.08}^{+0.20}$ & $<$615 & $>$0 & 5.2$_{-2.0}^{+2.7}$ & 2.4$_{-0.6}^{+3.5}$ & & 750$^{+230}_{-130}$ & 0.20$_{-0.10}^{+0.4}$ & $>$10 & 2000$_{-910}^{+3380}$ & 0.05$_{-0.02}^{+0.02}$ & 474/474\\
\vspace*{-2mm}\\
\hline
\vspace*{-2mm}\\
2008 &0.84$_{-0.03}^{+0.03}$ & 80$_{-10}^{+10}$ & 0.2 & 7.9$_{-0.4}^{+0.5}$ & 2.1$_{-0.2}^{+0.6}$ & 190$^{+20}_{-30}$ & 2.0 & $>$180 & 40$_{-20}^{+15}$ & 3.2 & 0.13 &525/515\\
%
%2008 &0.45$_{-0.02}^{+0.02}$ & $<$100 & $>$0 & 4.6$_{-0.2}^{+0.2}$ & 4.9$_{-0.1}^{+0.4}$ & & 380$^{+10}_{-10}$ & 2.3$_{-0.2}^{+0.2}$ & $>$150 & 20$_{-10}^{+5}$ & 9.92$_{-1.53}^{+5.49}$ & 572/515\\
\vspace*{-2mm}\\
\hline
\end{tabular}
\caption{\label{tabparam3} Results of the fits of the XIS and HXD data with a {\sc power law} (top table) or {\sc eqpair} (bottom table) for the continuum and the ionized reflection component {\sc reflionx}. {\bf In the case of the {\sc power law} fit, the photon index used in {\sc reflionx} is fixed to the one of the power law continuum. In the case of {\sc eqpair}, we froze the photon index  of {\sc reflionx} to the one obtained in the power law fit.}.  The inclination was fixed to 20 deg. We uses the kernel from the {\sc laor} line profile to account for the gravitational effects close to the black hole (model {\sc kdblur}) of {\sc xspec}). %. 
All errors are 90 per cent confidence for one parameter. The fluxes are in unit of ergs.cm$^{-2}$.s$^{-1}$ and are computed in the 0.7-70 keV energy range. The bolometric luminosity $L_{bol}$ is assumed to be the sum of the luminosities of the three spectral component {\sc diskbb, eqpair} and {\sc reflionx}. We adopt an Eddington luminosity of $L_{Edd}\simeq1.3\times 10^{39}$ erg s$^{-1}$ { (i.e. we assume a 10 solar masses black hole)} and a distance to GX 339-4 of 8 kpc.} %We have also reported the $\chi^2/dof$ corresponding to the XIS data alone, in the 0.7-10 keV range, in order to compare with the best fit results of Tab. \ref{tabparam}. }. 
\end{center}
\end{table*}

%\begin{table*}
%\begin{center}
%\begin{tabular}{cccccccccccccccc}
%\hline
%\multicolumn{15}{c}{{\it tbnew*(nthcomp+eqpair+kdblr*reflionx)}}\\
%Obs. & $N_h$ & $\Gamma_{nth}$ & $T_{e,nth}$ & $T_{bb,nth}$& $N_{nth}$ & $l_h/l_s$ & $\tau$ & $T_{e,eqp}$ & $T_{bb,eqp}$ & $N_{eqp}$ & $R_{in}$ & $\xi_{ref}$ & $N_{ref}$  & $\chi^2$/dof \\ %& $L_{1-100\ keV}^{unabs}/L_{Edd}$ 
%%\hline
% & $\times 10^{22}$ &  & $keV$ & $eV$ & $\times 10^{3}$ &  &  & $keV$ & $eV$ & $\times 10^{-11}$ & $R_{g}$ & & $10^{-6}$ &   \\ 
%%& $L_{1-100\ keV}^{unabs}/L_{Edd}$ 
%
%\hline
%\vspace*{-2mm}\\
%\vspace*{-2mm}\\
%\hline
%\end{tabular}
%\caption{  }. 
%\end{center}
%\end{table*}

%

\subsection{Broad band spectral analysis: ionized reflection}
\subsubsection{Ionized reflection}
\label{ionref}
From the previous sections, fits of the soft X-ray excess and iron line profiles, independently one with each other, both suggest an increase of the disk inner radius from OBS1 to OBS2. Its evolution is however uncertain for the three last observations. We also confirm the large value of $R_{in}$ deduced from the iron line fit in the 2008 Suzaku observation. To go a bit further, we use, in this section, more realistic models for the continuum and the reflection component. We take also advantage of the HXD instrument of Suzaku and include the HXD/PIN (20-70 keV) data in our fits.\\  

{We have reported In Fig. \ref{chi2contrib} the $\chi^2$ contribution when we extend the best fit model {\sc tbnew$\times$ (diskbb+ power law + laor)} of the XIS data (the best fit parameters are reported in Tab. \ref{tabparam2}) above 10 keV in the HXD/PIN  energy range. A clear excess between 20 and 40 keV is visible especially in OBS1 and OBS2 and suggests the presence of a reflection bump.}

To provide a more physical description of the reflection component giving birth to the iron line, we use the combination of the {\sc reflionx} code of \cite{ros05},  convolved with the relativistic kernel {\sc kdblur} \citep{lao91}. In agreement with the results of Sect. \ref{secge3kev}, we fix in {\sc kdblur} the outer disk radius $R_{out}$ to 400 $R_g$, the index of the radial power law dependence of the emissivity to 3 and the disk inclination angle to 20 degrees. %In {\sc reflionx}, for each observation the photon index is fixed to the value of the power law {\bf reported in Tab. \ref{tabparam2}}. The other parameters of {\sc reflionx} are the iron abundance, fixed here to 1, and the degree of ionization of the disk, $\xi_{ref}$. \\

{In {\sc reflionx} the illumination has a power law shape. So we first fit the different observations with a power-law for the primary continuum, {\bf  fixing the photon index in {\sc reflionx} to that of the power law continuum}. However, as already discussed in Sect. \ref{secxisge07}, comptonization of the soft disk photons in a hot corona is widely accepted as the mechanism at the origin of the X-ray continuum of X-ray binaries. Thus, instead of a simple power law, we also use the Comptonization model {\sc eqpair} \citep{cop99} even if it is not perfectly consistent with the use of {\sc reflionx}. {\bf In this case, we fix the photon index in {\sc reflionx} to the precedent values obtained when fitting with the power law continuum \footnote{\bf The slope of {\sc eqpair} can be estimated from the $l_h/l_s$ ratio (e.g. \citealt{mal01}) and appears to be very close to the photon index of the power law fits.}} We will see that our results are qualitatively similar between the two models, the advantage of  {\sc eqpair} being that its main parameters (i.e. the ratio between the compactness of seed photons, $l_s$, and hot electrons, $l_h$, the corona optical depth $\tau$ and the temperature of the soft disk photons $T_{bb}$) have a direct physical meaning.}\\

The best fit parameters obtained with these models are reported in Tab. \ref{tabparam3}. { Note that the fits are always better with {\sc eqpair} than with a power law. The lack of a low energy roll-over in this latter case could explain the larger residuals observed in the soft  band ($<$ 2 keV). The best fit parameters are however qualitatively similar between the two models, the main difference being the reflection component in OBS1 which has a lower flux and lower ionization with the use of a power law continuum. But $R_{in}$ is still well constrained to a small value $\sim$ 15 $R_g$ in this observation. Note also that the ionization parameter is not larger in OBS3 compared to the other observations and then does not support the presence of a more ionized iron line as suggested by the fits between 3 and 10 keV with a simple gaussian (see Sect. \ref{secge3kev}).} \\

Since both models (with a power law or {\sc eqpair} for the continuum) give similar parameter constraints and since the fits with {\sc eqpair} gives always a better $\chi^2$, from now on we will only discuss the results obtained with this model.
We find good fits in all cases but OBS1 for which some features are still present in the soft energy range ($<$ 1 keV, see next section). The unfolded spectra as well as the data/model ratios of each observation are reported in Fig. \ref{fiteqpair}. In order to make a direct comparison to the {\sc diskbb + power law + gaussian} fits below 10 keV discussed in Sect. \ref{secxisge07} (and reported in Tab. \ref{tabparam2}), we have computed the corresponding $\chi^2/dof$ of these new fits but limited to the 0.7-10 keV energy range. We find the following values: 530/382, 401/382, 392/379, 417/372 and 355/341 for the five (from OBS1 to OBS5) observations respectively. The improvement of the fits in the 0.7-10 keV energy range with the blurred ionized reflection model are really significant for OBS1 and OBS2  with $\Delta\chi^2=$ 49 and 27 for two less degrees of freedom. No significant improvement is obtained for the other observations, potentially because of their lower statistics. \\

The hard to soft compactnesses ratio $l_h/l_s$ increases smoothly all along the monitoring in agreement with the observed spectral hardening (see Tab. \ref{tabparam}). It is of the order of 5-8 suggesting a photon starved geometry for the hot corona. On the other hand, the hot corona optical depth, $\tau$, increases significantly between OBS1 and OBS2 by a factor $\sim$4 and then stays roughly constant of the order of $\sim$ 2-3. Concerning the ionization parameter, apart from OBS2, it is relatively high (with admittedly large error bars) of the order of $\xi_{ref}\sim$1000, suggesting a ionized reflecting medium.\\

\begin{figure*}
\begin{tabular}{cc}
\includegraphics[width=\columnwidth]{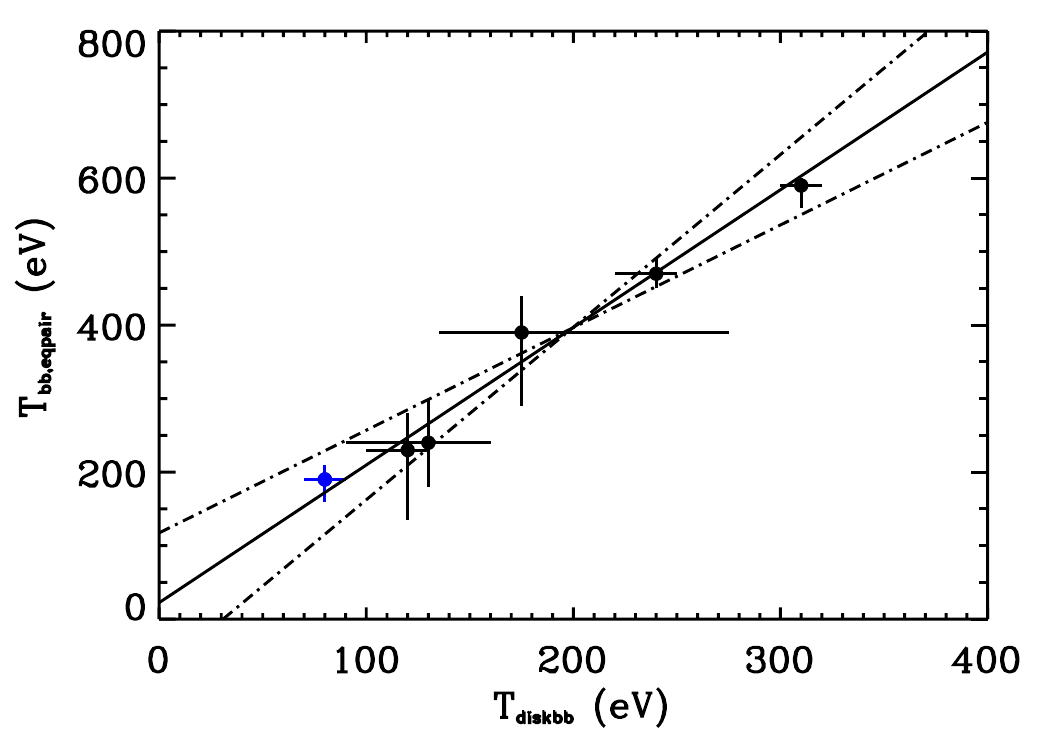} & \includegraphics[width=\columnwidth]{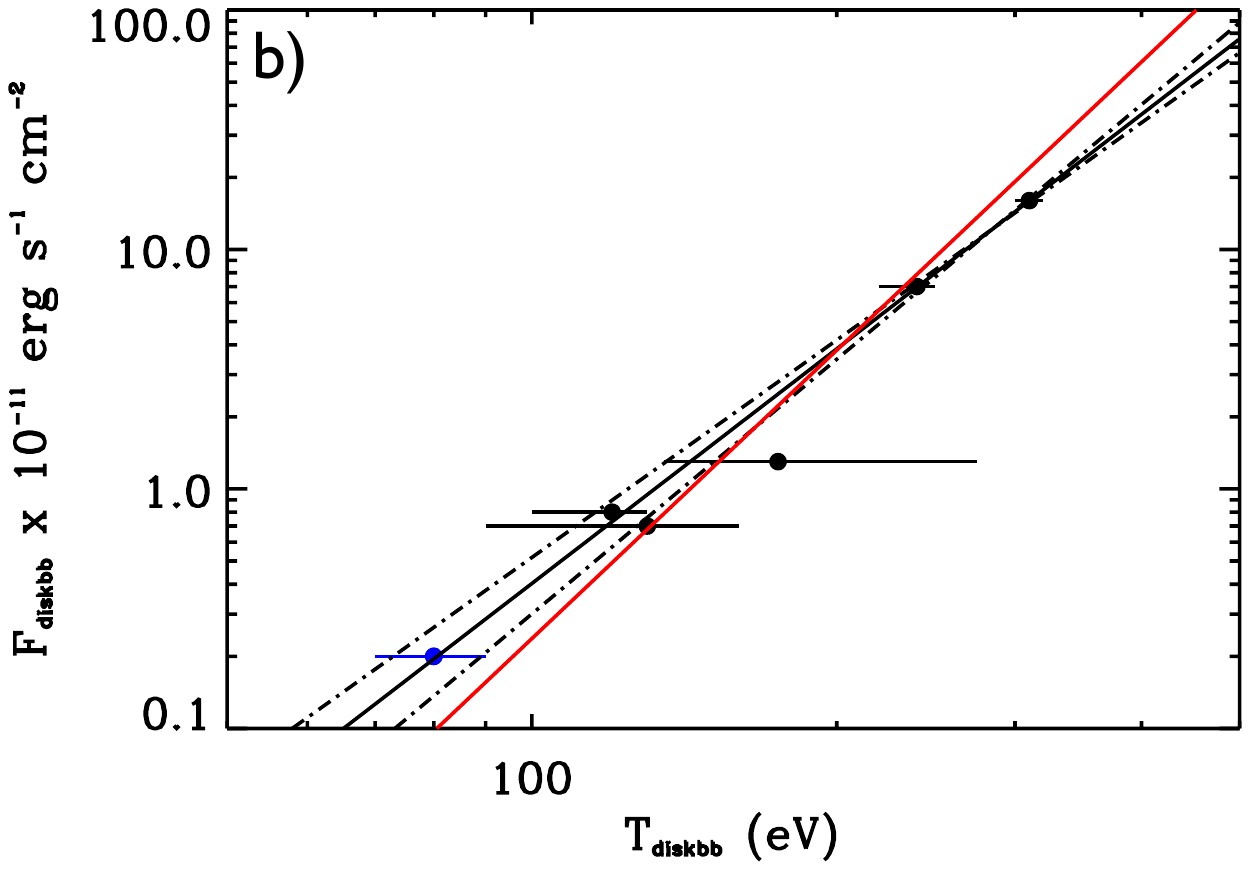}\\
\includegraphics[width=\columnwidth]{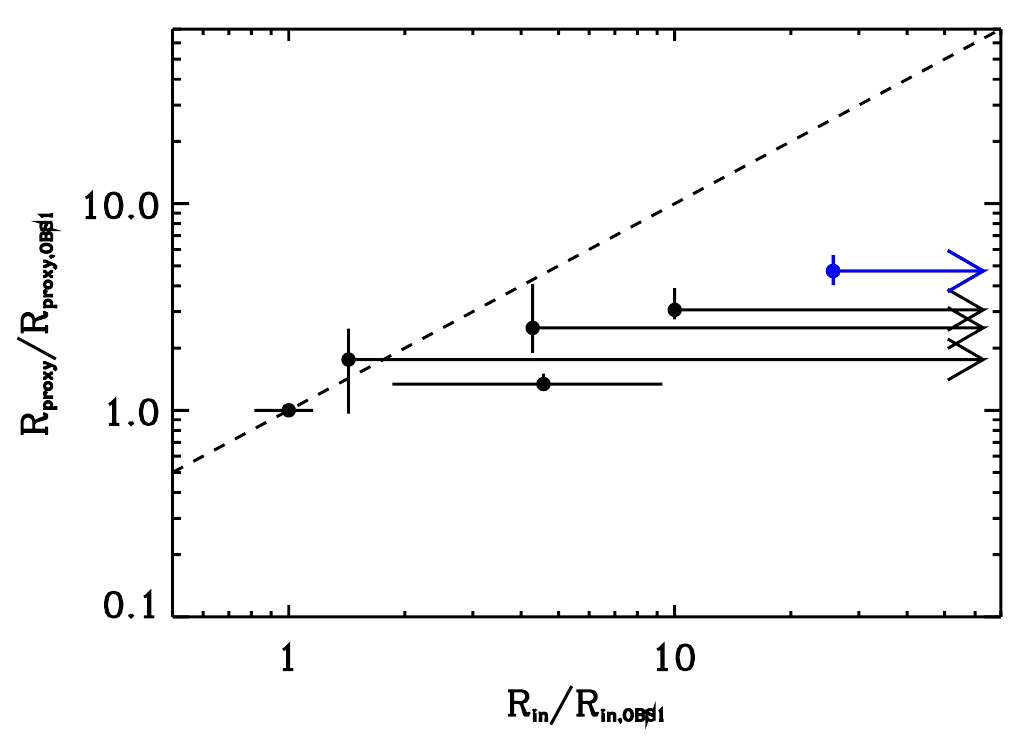} & 
%\includegraphics[width=0.5\columnwidth]{tbbdisk-tbbeqpair.pdf} & \includegraphics[width=0.5\columnwidth]{fdiskbb-tbbdisk.pdf}\\
%\includegraphics[width=0.5\columnwidth]{rproxy-rfit.pdf} & 
%\end{tabular}
%\begin{minipage}{1.\columnwidth}
%\vspace*{-6.9cm}
\begin{minipage}{\columnwidth}
\vspace*{-6.9cm}
 \caption{{\bf a)} Best fit {\sc diskbb}  temperature vs {\sc eqpair} soft photon temperature. The solid line is the linear best fit. It as a slope of 1.9. The dot-dashed lines corresponds to the 1$\sigma$ error on the slope i.e. 1.9$\pm$0.5. {\bf b)} Flux of the {\sc diskbb} component versus {\sc diskbb} temperature. The black solid line corresponds to the log-log linear best fit $\displaystyle F_{diskbb}\propto T_{diskbb}^{3.3}$ and the dot-dashed lines to the 1$\sigma$. The red solid line correspond to the best fit assuming a $T_{diskbb}^4$ law. {\bf c)} Plots of $R_{proxy}$ (see Sect. \ref{secrec}) versus the disk inner radius $R_{in}$ deduced from our fits (reported in Tab. \ref{tabparam3}) and assuming the X-ray luminosity $\propto\dot{M}^{\alpha}$ with $\alpha$=2.5 { (the results are very similar for $\alpha$=2 or 3)}. The dashed line corresponds to $R_{proxy}=R_{in}$. The blue points in each figure correspond to the 2008 Suzaku observation.}
\label{resfiteq}
\end{minipage}\\
\end{tabular}
\end{figure*}

The 0.7-70 keV fluxes of the different spectral components (respectively $F_{diskbb}$, {$F_{eqpair}$ and $F_{ref}$) are also reported in Tab. \ref{tabparam3}. They show a clear decrease along the monitoring in agreement with the fact that the source returns back to its quiescent state. However, the relative ratios $F_{diskbb}/F_{eqpair}$ and $F_{ref}/F_{eqpair}$ are not constant, as we could expect if all the spectral components fade in the same way. These ratios decrease from $\sim$14 to 1\% and $\sim$25 to 10\% respectively. \\ 

The {\sc diskbb} temperature, $T_{diskbb}$, as well as the {\sc eqpair} soft photon temperature, $T_{bb}$, show also clear decrease from OBS1 to OBS5. Interestingly, while the two temperatures were let free to vary during the fitting procedure, {$T_{diskbb}$} is about half {$T_{bb}$} (see Fig. \ref{resfiteq}a). Such correlation may have a physical origin. Indeed, in {\sc eqpair} the soft temperature $T_{bb}$  corresponds to the disk temperature actually "see" by the hot corona. On the other hand, $T_{diskbb}$ is the effective disk temperature. The ratio between the two, the so-called spectral hardening factor, is generally estimated to be of the order of 2-3 (e.g. \citealt{shi95,sob99,mer00,dav06}) in agreement with what we find.\\

For comparison we have also fitted the 2008 Suzaku observation with the same model. The best fit parameters are reported in the last row of Tab. \ref{tabparam3}. Interestingly, apart from the reflection ionization parameter which is of the order of $\sim$40 i.e. well below our best fit values observed in 201 and a larger disk inner radius, the other parameters agree well with the spectral evolution from OBS1 to OBS5. More precisely, and even if there is a 3 year gap between them as well as a poorer statistics in 2011, the 2008 observation seems to be very similar to OBS5 but with a different accretion disk state.\\

\begin{figure*}[!th]
\begin{tabular}{cc}
\includegraphics[width=\columnwidth]{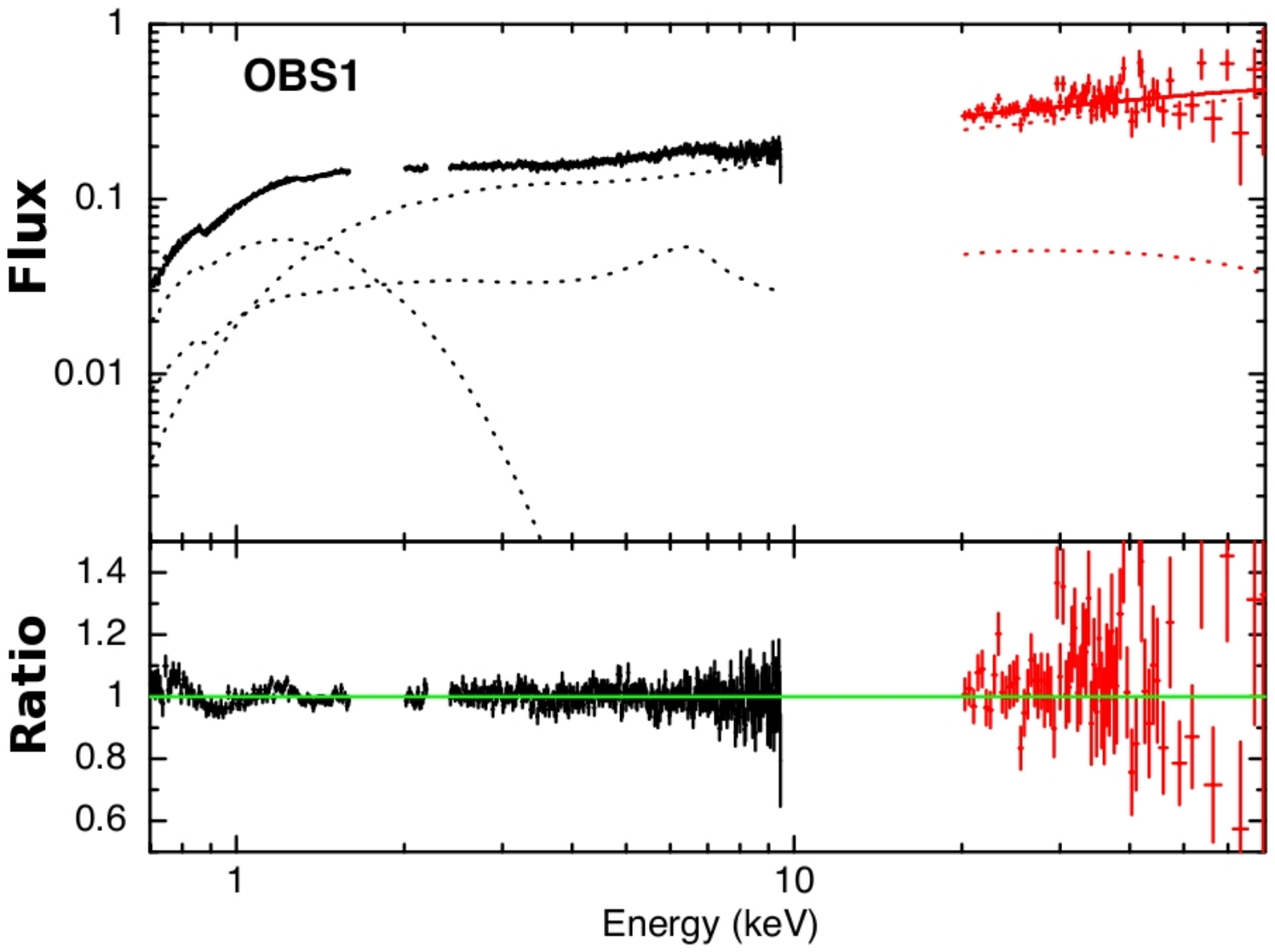} & \includegraphics[width=\columnwidth]{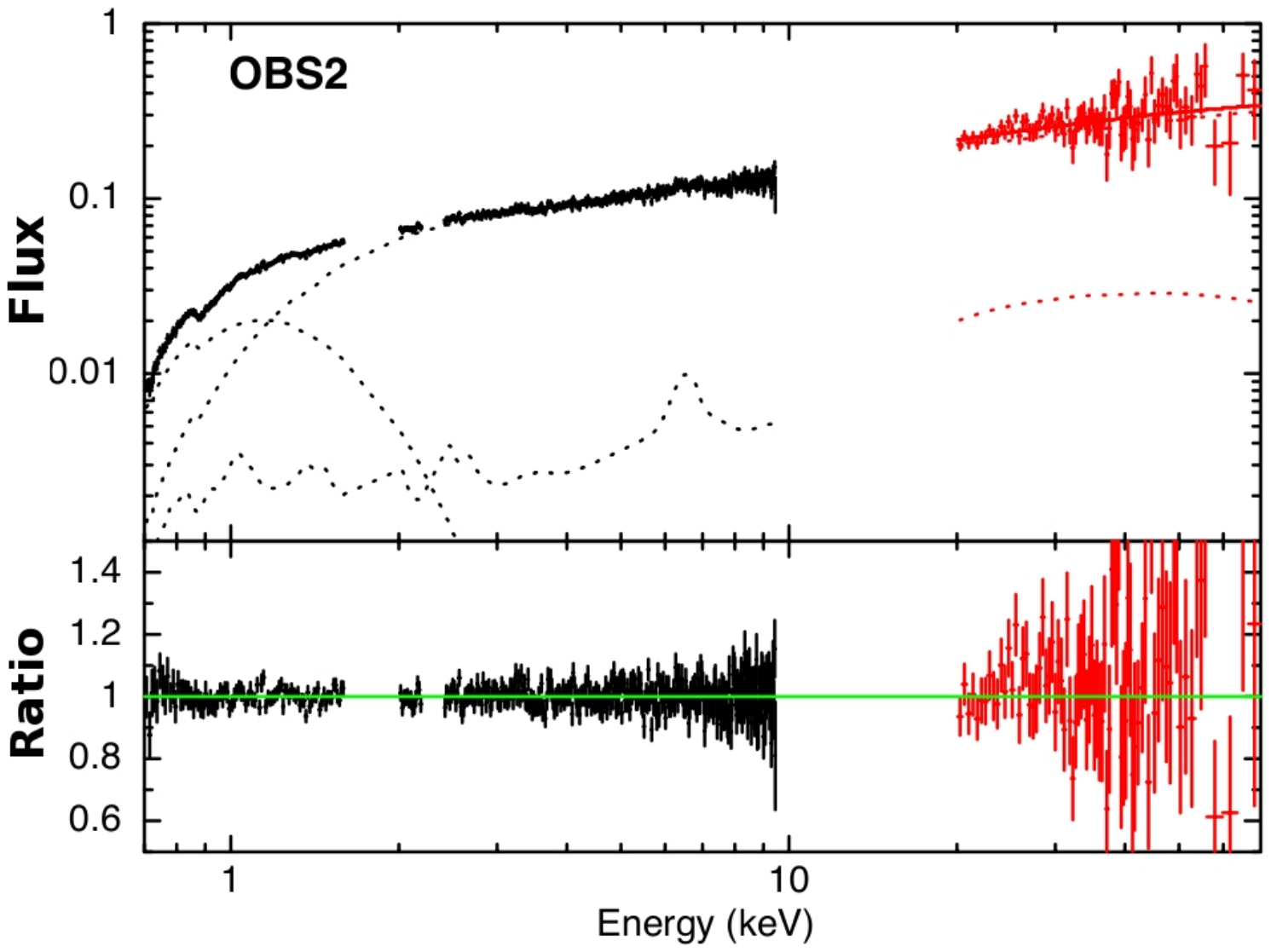}\\
\includegraphics[width=\columnwidth]{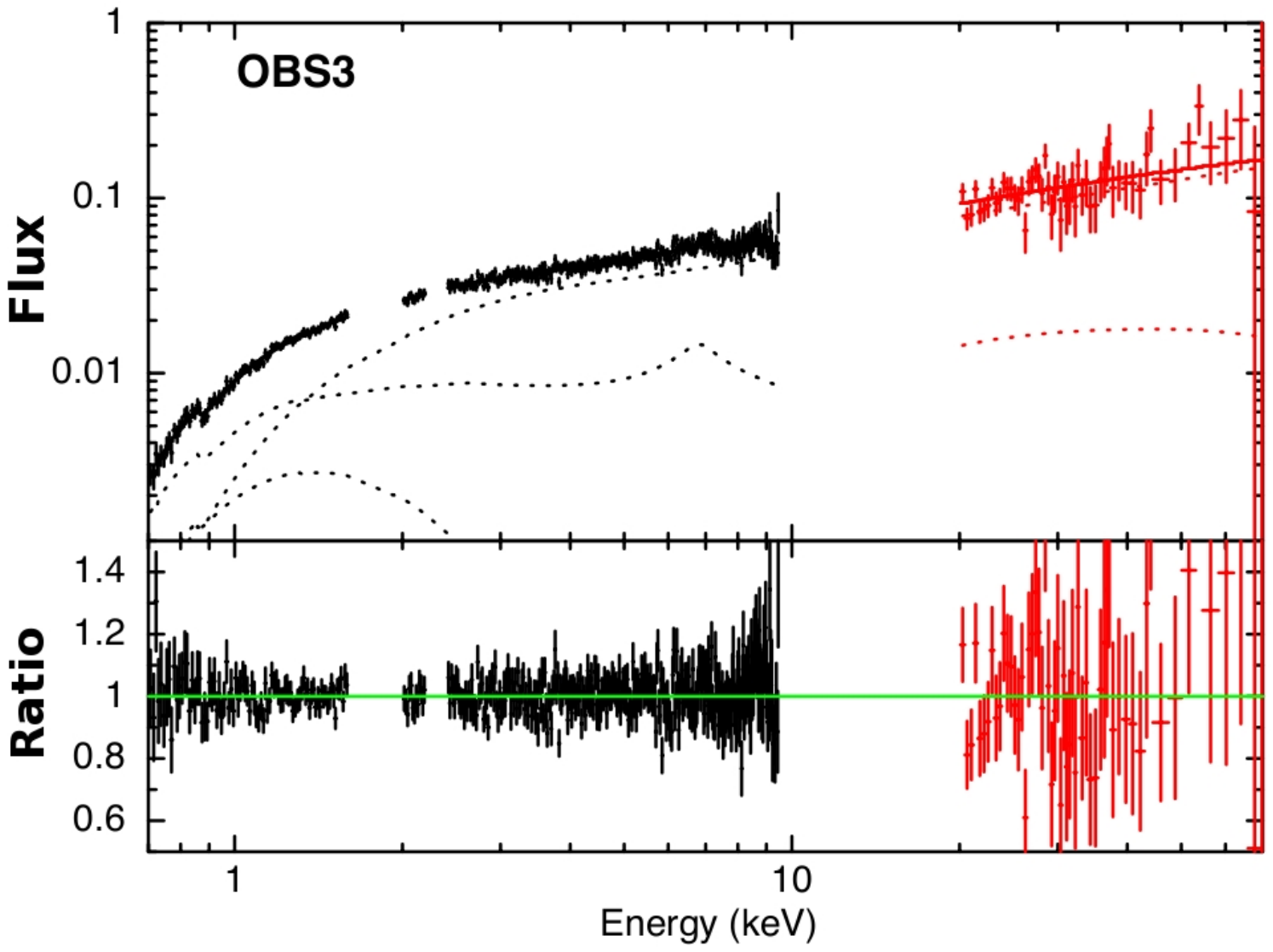} & \includegraphics[width=\columnwidth]{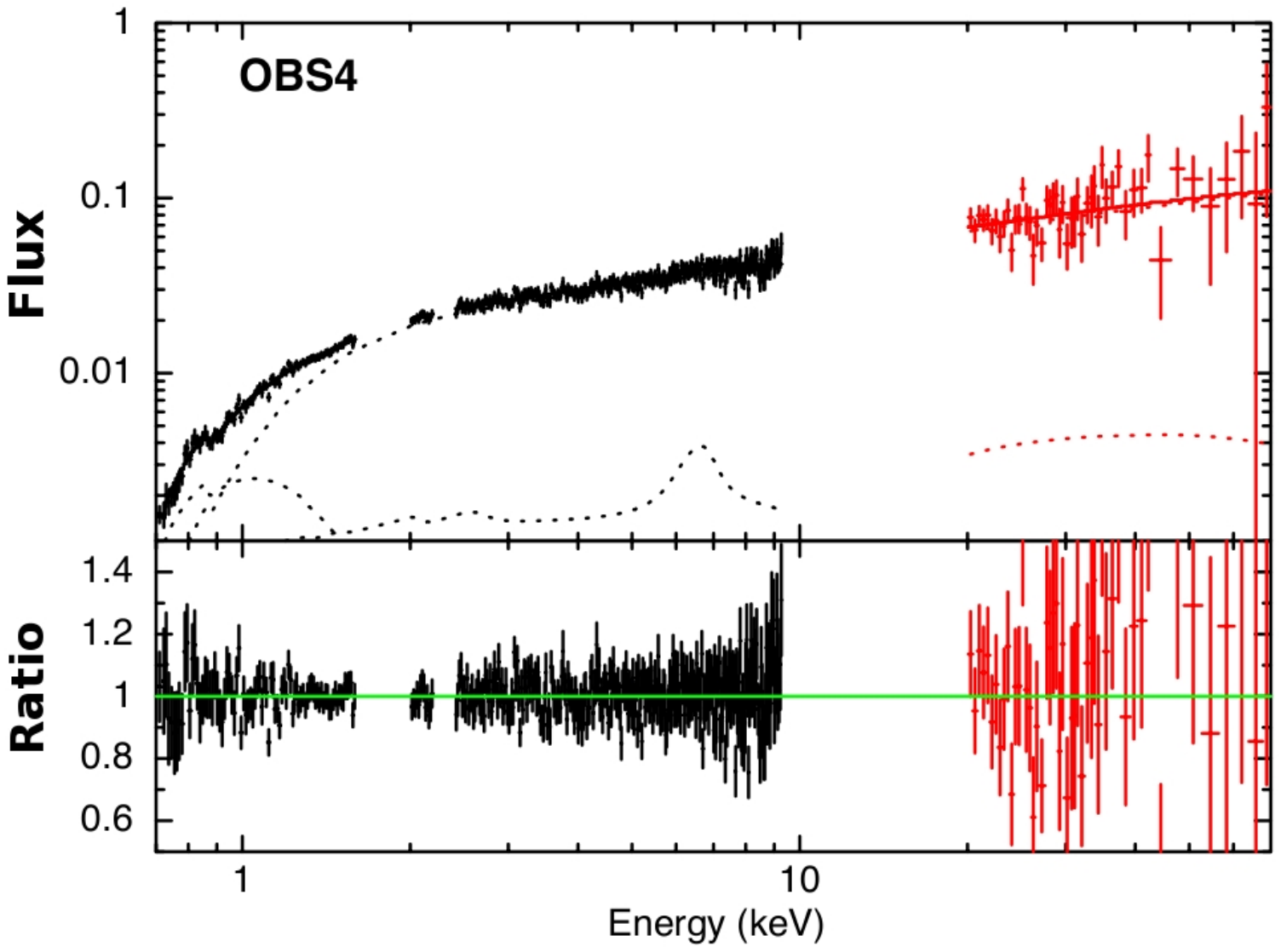}\\
\includegraphics[width=\columnwidth]{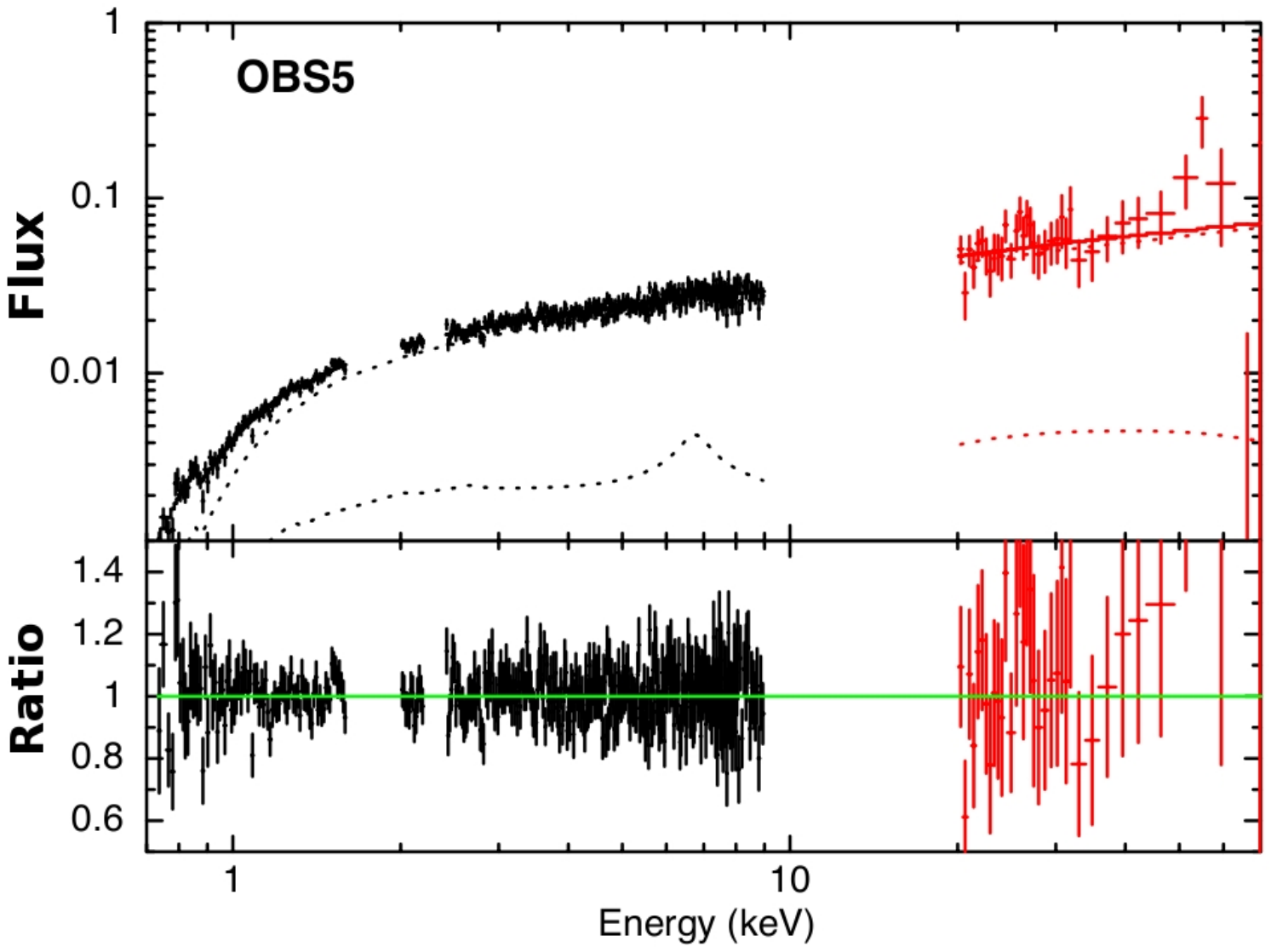} & \includegraphics[width=\columnwidth]{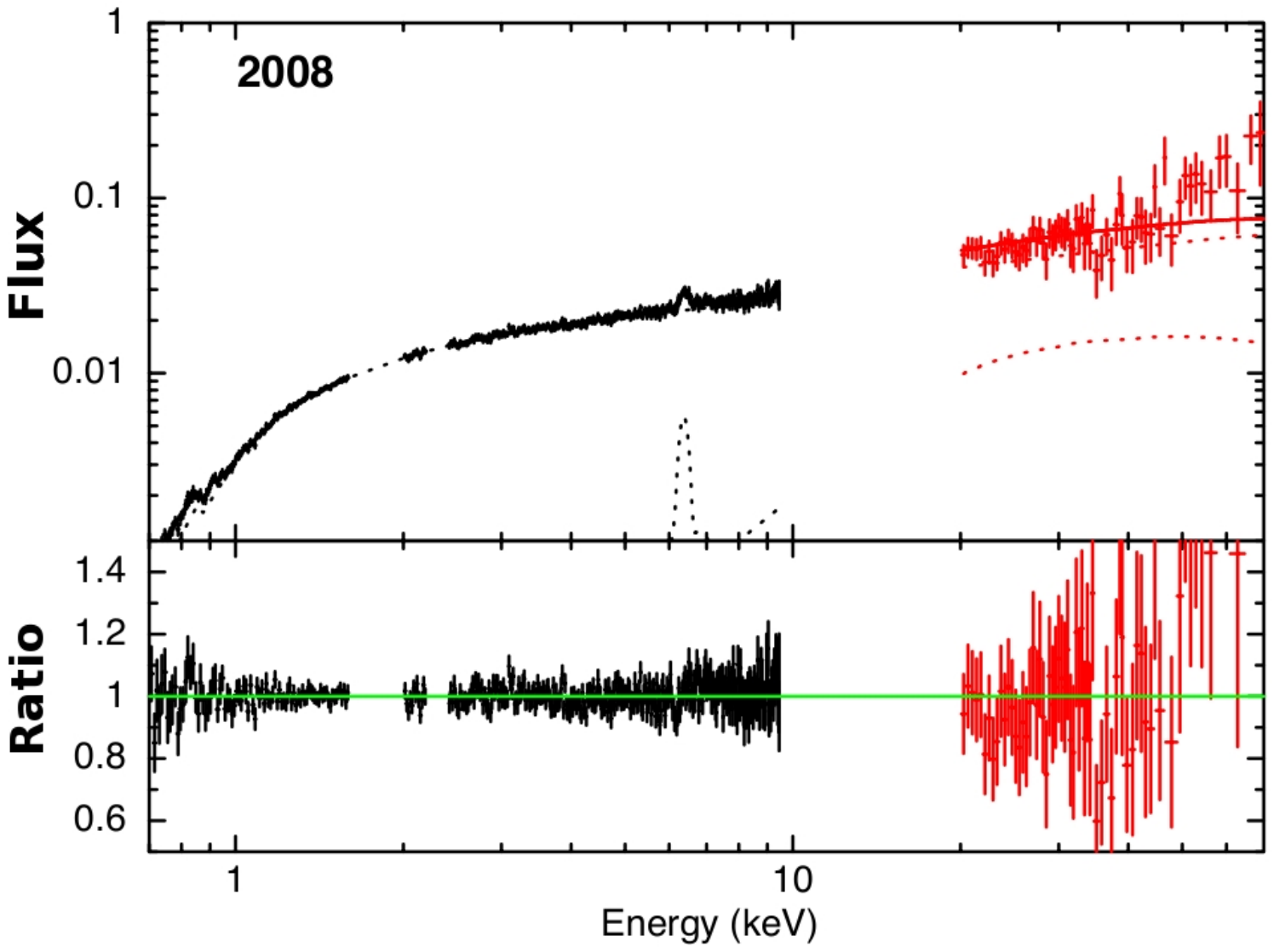}
\end{tabular}
%\begin{minipage}{\columnwidth}
%\vspace*{-5cm}
 \caption{Unfolded best-fit model of the five Suzaku observations (XIS: black crosses; HXD/PIN: red crosses) and corresponding data/model ratios using {\sc eqppair} for the continuum, a blurred reflection {\sc kdblur$\otimes$reflionx} for the reflection component and a multicolor disk {\sc diskbb} for the soft X-rays.}
\label{fiteqpair}
%\end{minipage}
%\end{tabular}
\end{figure*}

We have checked if the uncertainties on $R_{in}$ in OBS5 could be due to the lower statistics compared to 2008 by simulating a set of data from the best fit  model obtained for 2008 but with a combined XIS0+XIS3 exposure time of 40 ks (i.e. of the order of the exposure time of OBS5) instead of 210 ks. A fit of this simulated spectrum gives $R_{in}>130 R_g$ and $\xi_{ref}=70_{-20}^{+60}$. So if the OBS5 spectrum and line shape were exactly the same as in 2008, we should have obtained values of $R_{in}$ and $\xi_{ref}$ in agreement with those observed in 2008, even if the statistics of OBS5 is low. This is in contradiction with our results and this suggests that the accretion disk is in an intrinsically different state between 2008 and 2011.

\subsubsection{Ionized absorption}
{ As said previously, the best fit of OBS1 is not very satisfactory and features are visible, especially in the soft X-ray range, in the data/model ratio reported in Fig. \ref{fiteqpair}. The presence of absorption lines in the soft X-ray spectrum of GX 339-4 have already been reported in the literature when the source was in an intermediate state transiting to the low/hard state (e.g. M04, \citealt{jue06}). If most of these lines may be produced by the Interstellar Medium \citep{jue06}, a sizable contribution of a few of them (like e.g. the Ne~{\sc ix} line at 0.922 keV) could be due to a local contribution from an intrinsic AGN-like warm-absorber perhaps produced by a disk wind.}\\

We have tested the presence of such absorber by adding an ionized absorber component (model {\sc absori} of {\sc xspec}) in our fits. The fit of OBS1 improves significantly ($\Delta\chi^2$=102!) thanks to the addition of this spectral component. The corresponding parameters of the absorber are reported in Tab. \ref{tabparam4} i.e. an hydrogen column density of $N_{h,abs}\sim10^{22}$cm$^{-2}$ and an ionization parameter $\xi_{abs}\sim$300. The other model parameters do not significantly change compared to Tab. \ref{tabparam3}.
We add the same {\sc absori} component to the other observations (including 2008). We do not find any significant improvement. The best fit are also reported in Tab. \ref{tabparam4}. The decrease of the signal-to-noise ratio may however limit the correct detection of absorbing material \footnote{\bf Note that, despite the difference in the the soft band due to the lack of a low energy roll-over when fitting the continuum with a power law, we obtain consistent values of the absorber parameters when we use this model instead of {\sc eqpair}}. \\ 

{ Then, we can have a rough estimate of the maximal distance $d$ between the absorbing material and the X-ray source by assuming that $d$ is necessarily larger than the radial extension $\Delta r$ of the absorber. Since $N_{h,abs}=n\Delta r$ (n being the density of the warm absorber) and $\xi_{abs}=\displaystyle\frac{L_{bol}}{d^2n}$ then $d>\Delta r$ implies $d<\displaystyle\frac{L_{bol}}{\xi_{abs}N_{h,abs}}$. Using the best fit parameter values obtained for OBS1 i.e. $L_{bol}\simeq 10^{37}$ erg s$^{-1}$, $\xi_{abs}\simeq$300 erg cm s$^{-1}$  and $N_{h,abs}\simeq10^{22}$cm$^{-2}$ we find that the absorber should be at a maximal distance of $\sim 10^6 R_g$ (assuming a 10 black hole solar masses) from the central X-ray source. {\bf This is of the order of the binary separation in GX 339-4 as estimated by \cite{zdz04b}, thus agreeing with the fact that the absorber may be produced in the inner parts of the binary system.}}\\

{ To have a qualitative idea of the ions potentially responsible for the observed absorption features in OBS1, instead of {\sc absori} we simply add gaussian absorption lines, with width fixed to 0 eV (a more detailed work on these absorption line is dedicated to a future work). We focus only on OBS1 since the other observations do not need apparently the addition of absorption components. As previously said, a few absorption lines were already observed in GX 339-4 in past Chandra observations (M04), the most intense one being Ne~{\sc ix} at 0.922 keV. We thus add an absorption gaussian line with energy in the range 0.9-0.95 keV. The fit improves strongly with $\Delta\chi^2$=61 for 3 less degrees of freedom. The gaussian best fit energy $E_{abs}$=0.92$\pm$0.02 is consistent with Ne~{\sc ix}\footnote{{ Note that this absorption line energy is well below the Si edge  which is known to contaminate the XIS response around 1.8 keV}}. The line has an equivalent width (EW) of $\sim$5.7 eV which corresponds to a Neon column density in between $5\times 10^{16}$ and $10^{17}$ cm$^{-2}$ (see M04 for the computation of the Neon column density). This agrees completely with the estimates measured by M04 and \cite{jue06} and this is still higher than the expected value from the ISM, suggesting also an origin in the local environment of the X-ray binary. {\bf  \cite{luo14} reach the same conclusion for GX 339-4 as well as for a sample of 11 other X-ray binaries}. We have tested that the addition of such gaussian absorption line is not needed in the other observations.\\

We have also tested the presence of (weak) blue-shifted ionized absorption lines from Fe {\sc XXV} and Fe {\sc XXVI} since they have been observed in a few microquasars and interpreted as signature of fast ionized outflows. Their expected EW are generally of the order of 10-20 eV (e.g. \citealt{pon12}).  We do not detect such components is our data and find only upper limit for their equivalent width of 5-10 eV.}}

\label{sectionized}
\begin{table}
\begin{center}
\begin{tabular}{ccccc}
\hline
\multicolumn{5}{c}{{\it Ionized absorber}}\\
Obs. & $N_{h}$ & $\xi_{abs}$ & $\Delta\chi^2$ & $\chi^2/dof$ \\
 & $\times 10^{22}$ & & \\
\hline
\vspace*{-2mm}\\
1 & 1.0$^{+0.2}_{-0.2}$ & 320$^{+90}_{-60}$ & 107 & 597/513\\
2 & 0.4$^{+0.1}_{-0.3}$ & 180$^{+230}_{-130}$ & 3 & 541/513\\
3 & $<$4.0 & $>90$ & -1 & 520/510\\
4 & $<$2.4 & $[0-5000]$ & 1 & 551/503\\
5 & 7.4$^{+3.2}_{-3.7}$ & $>$4900 & 3 & 472/472\\
\vspace*{-2mm}\\
\hline
\vspace*{-2mm}\\
2008 & $<$0.25 & $[0-5000]$ & 0 & 524/513\\
\vspace*{-2mm}\\
\hline
\end{tabular}
\caption{\label{tabparam4} Best fit parameters of the ionized absorber component. The $\Delta\chi^2$ is the $\chi^2$ variation compared to the best fits reported in Tab. \ref{tabparam3}}. 
\end{center}
\end{table}

%In the case of OBS3, 4 and 5, the best fits may also be outside the validity range of {\sc refbhb} especially concerning the disk temperature which is expected to decrease below 0.15 keV, the lower limit of the table used for {\sc refbhb} (cf. Fig. \ref{compardincer}). This could explain why the fits pegged to the maximal values of $H_{den}$ and $F_{Illum}/F_{BB}$ (see Tab. \ref{}).

\section{Hints of disk recession}
\label{secrec}
To be correct, when we talk about disk recession, we are talking about the recession of the optically thick part of the accretion flow. It is possible that the accretion disk extends down to the inner more stable orbit but that, for different reasons (e.g. most of its accretion power is advected or ejected), it does not radiate any more.\\

While we do not find clear indications of disk recession during our monitoring, a few arguments are consistent with this interpretation.
First, and contrary to fits below 10 keV, fits of the broad band spectra with blurred ionized reflection give good constrains on $R_{in}$ for OBS1 and OBS2, of the order of 10 $R_g$ and $\sim$30$R_g$ respectively, but it also puts lower limits  for OBS3 ($> 70 R_g$), OBS4 ($>10 R_g$) and OBS5 ($>30 R_g$).  For the 2008 data set, our fit gives a lower limit $R_{in}>180 R_g$, in agreement with T09 and clearly suggesting a disk recession {\bf (but see \cite{fab14} for the limitations of the use of X-ray reflection to estimates  the disk inner radius)}. The fact that this recession was apparently stronger in 2008 could be due to the fact that the source was in a persistently faint hard state since months. This could have let enough time for the inner accretion flow to evolve into the observed configuration.\\

Our estimates of $l_h/l_s$ from our fits also agree with the work of \cite{sob11}. { Note however that we are not using exactly the same model as these authors. No reflection component was taken into account in their work but an iron line. We do not expect however that this would have any strong effect on our estimate of $l_h/l_s$ compared to their methods given the low flux we found in the reflection component (i.e. a factor 10 below the continuum flux, see Tab. \ref{tabparam3}). Thus we believe that our results can be safely compared to theirs}. These authors studied the changes of the $l_h/l_s$ ratio with the bolometric luminosity $L_{bol}$ in the hard states of GX 339-4 and GRO J1655-40. At luminosities of the order of $\sim$ 0.1-1\%, i.e. the range of luminosities of the present Suzaku observations, \cite{sob11} confirmed a transition in behaviour of $l_h/l_s$  with $L_{bol}$. At luminosities lower than $\sim$ 0.1-1\%, both parameters correlate while they anticorrelate at larger $L_{bol}$. According to these authors, this behavior is consistent with a scenario where seed photons change from cyclo-synchrotron, at the lowest luminosities, to those from a (truncated) disk, at higher luminosities. Our observed increase of $l_h/l_s$ from OBS1 to OBS5, i.e. with a decrease of $L_{bol}$, suggests then that the emission of the accretion disk is still dominating the cooling process of the hot corona.\\

But then, the changes of $l_h/l_s$ implies a variation of the disk-corona geometry. The increase of $l_h/l_s$ from OBS1 to OBS5  indicates  a faster decrease of the soft photon flux compared to the corona heating power. This situation is naturally expected if  $R_{in}$ increases from OBS1 to OBS5. Note that the increase of $R_{in}$ would also imply a diminution of the reflecting area and consequently a faster decrease of the reflecting component compared to primary one. This is also in agreement with the decrease of $F_{ref}/F_{eqpair}$ as discussed in Sect. \ref{ionref}. \\

We have reported in Fig. \ref{resfiteq}b the flux $F_{diskbb}$ of the {\sc diskbb} component versus the {\sc diskbb} temperature $T_{diskbb}$. The log-log linear best fit (black solid line in Fig. \ref{resfiteq}b) gives $\displaystyle F_{diskbb}\propto T_{{diskbb}}^{3.3\pm0.2}$ i.e. a bit smoother than the Stefan-Boltzmann law in $T^{4}$ expected in the case of an optically thick accretion disk with fixed inner radius. This discrepancy could be explained by a variation of the hardening factor with the disk temperature (e.g. \citealt{sal13}). However, we can again interpret this result in term of variation of the disk inner radius. Indeed,  in an standard accretion disk of inner radius $R_{in}$, inner temperature $T_{diskbb}$ and accretion rate $\dot{M}$ we expect:
\begin{equation}
T_{diskbb}^4R_{in}^3\propto\dot{M}.\label{eqrproxy}
\end{equation}
On the other hand, at the end of the outbursts, X-ray binaries are known to be in a radiatively inefficient state where the X-ray luminosity is proportional to $\dot{M}^{\alpha}$ with $\alpha\sim 2-3$ (e.g. \citealt{cori11} an references therein). Then $F_{eqpair}^{1/\alpha}$ can be used as a proxy for the accretion rate of our system. In consequence, the ratio $R_{proxy}=[F_{eqpair}^{1/\alpha}/T_{diskbb}^4]^{\frac{1}{3}}$ should be a proxy of the disk inner radius $R_{in}$.  We have plotted $R_{proxy}$ normalized to its value in OBS1 (and assuming $\alpha=2.5$) in  Fig. \ref{resfiteq}c  versus the best fit inner radius values $R_{in}$ obtained from our broad band fits.  While the values of $R_{proxy}$ for the last 4 observations are always larger than the one computed for OBS1, and then supporting the scenario of the recession of the optically thick part of the accretion disk, they are smaller than $R_{in}$ in most cases. This discrepancy may be due to the bad estimate of the disk inner temperature $T_{diskbb}$ from our fits due to the limited energy range of Suzaku in the soft X.\\

\section{Disk-wind to disk-jet transitions?}
\label{diskwind}
In the recent context of outflow/wind signatures detection in black hole X-ray binaries (e.g. \citealt{mil06c,pon12,dia12}) and their potential link to the jet evolution during outburst \citep{mil06d,nei09}, the absorption features significantly detected in OBS1 (and only in OBS1) is an interesting results. Let's recall that the re-ignition of the radio emission was detected just before the beginning of our campaign, peaking between OBS1 and OBS2, and interpreted  as the beginning of the jet-structure re-building  while the source turned back to the hard state (e.g. C13). By analogy with the observation of the evolution between a jet-dominated to a wind-dominated accretion flow  during a hard-to-soft transition  in GRS 1915+105 \citep{nei09}, our results could correspond to the reverse situation i.e. the evolution from a wind-dominated to a jet-dominated accretion flow during a soft-to-hard transition. 

{ The absorption features in OBS1 agree with the presence of Neon absorption lines, especially from Ne~{\sc ix} and its column density suggests that part of this line could be produced locally. %The former agree with a ISM origin while a large part of the latter could be produced locally. 
Interestingly, this line was detected in a Chandra observation of GX 339-4 when the source was in an intermediate state transiting to the low/hard state while OBS1 was observed just before the complete state transition back to the hard state. The disappearance of the Ne~{\sc ix} in the other observations of the monitoring could suggest that it is linked to a disk wind only present before the transition to the hard state. The absence of detection of Fe~{\sc xxv} and Fe~{\sc xxvi} however seems to indicate that this wind, if really present in OBS1, is not highly ionized.} %outflows however. But such outflows are expected to be mainly equatorial (e.g. \citealt{bel83,mel91,prog02,luk10} and may not be detected if the source is not highly inclined with respect to the line of sight \citep{pon12}.}

%COMPARER SIGMAT4 ET FAIRE RAPPORT ENTRE OBS5 ET 2008
%DIRE QUE MODELE ASSEZ DEGENERE DONC SOLUTION MONTREE P€S FORCEMENT UNIQUE

%DIRE QUE NH PEUT ETRE LIE A LA PRESENCE D'UN WIND!!!!

%%%%%%%%%%%%%%
\section{Concluding remarks}
%%%%%%%%%%%%%%
\label{discuss} 
Our Suzaku monitoring of GX 339-4 at the end of its 2010-2011 outburst caught the source during its soft-to-hard state transition. Simultaneous radio-OIR observations showed the recovery of the radio emission and were interpreted as signature of the re-ignition of the compact jets (C13). The onset of the radio emission occured between OBS1 and OBS2 while the re-flare observed in OIR reached its maximum close to OBS3 and OBS4.\\

The Suzaku observations show a global fading of the X-ray flux during the monitoring. For the spectral fits, we first use phenomenological models with a simple power law for the continuum. The addition of an iron line is statistically needed in all the observations. Fits with a {\sc laor} profile are statistically equivalent to fits with a gaussian. Simultaneous fits of the 5 observations with a {\sc laor} profile give a disk inclination angle of $\sim$ 20 deg. The constraints on the inner radius agree with $R_{in} < 10 R_g$ in OBS1 and OBS2 at 90\% confidence level. Due to the lower statistics, these constraints becomes $< 100 R_g$, for the 3 last observations.\\

The presence of a soft X-ray excess, above the 3-10 keV power law extrapolation, is clearly visible in the two first observations and the addition of a multicolor disk component improves statistically the fits. This is not the case for the three last observations. The evolution of the disk component (decrease of its inner temperature and total flux)  agrees with a disk recession from OBS1 ($R_{in}\sim$ 10 $R_g$) to OBS2 ($R_{in}\sim$ 20 $R_g$). However, again, due to the low statistics, we cannot confirm this recession for the last observations. A comparison with the very long Suzaku observation of GX 339-4 during a long faint hard state of its 2007-2008 outburst (T09), in a flux state similar to our OBS5, shows a relatively good agreement of the spectral shape between the two observations. The line shape is however inconsistent between the two pointings. In 2008 the source was observed during a persistent, but faint, hard state while in 2011 it was clearly in the decreasing phase of the outburst. It is then possible that the accretion flow had more time to evolve into a truncated accretion disk in 2008 compared to 2011.\\

The use of a model including blurred ionized reflection and thermal comptonisation continuum gives better fits than the simple {\sc power law + diskbb} model used previously, at least for OBS1 and OBS2. For the other observations both models give similar results. This model give constraints on $R_{in}$ in marginal agreement with a disk recession. Hints of such recession come also from the increase of the $l_h/l_s$ compacity ratio of the hot corona all along the monitoring, and from the deviation of the disk flux to the Stefan law in $T^4$. These hints have to be taken with care however given the low statistics and the complexity of the model that may lead to strong degeneracy between parameters.\\

Finally, signatures of ionized absorption seem to be present at least in OBS1 but absent in the other observations. The radio re-ignition occurring in between OBS1 and OBS2 (see C13),  we suggest that, during these two observations, the accretion flow may have transited from a disk wind, an ubiquitous characteristic of soft states, and a jet, signature of the hard states. These absorption features may be the last signature of  the disk wind before the transition to a jet-dominated state. \\

\section*{Acknowledgments}
The authors thanks the referee for a careful reading of the manuscript and for his/her comments that well improve its quality. POP acknowledges financial support from CNES. This work is part of the CHAOS project ANR-12-BS05-0009 supported by the french Research National Agency (http://www.chaos-project.fr)Ó.
This research has made use of data obtained from the Suzaku satellite, a collaborative mission between the space agencies of Japan (JAXA) and the USA (NASA).

%\bibliographystyle{/Users/petrucpi/Library/texmf/tex/latex/aa-package/bibtex/aa}
%\bibliography{/Users/petrucpi/Boulot/COM/BIBLIO/ref}

\begin{thebibliography}{68}
\expandafter\ifx\csname natexlab\endcsname\relax\def\natexlab#1{#1}\fi

\bibitem[{{Barret} {et~al.}(2000){Barret}, {Olive}, {Boirin}, {Done},
  {Skinner}, \& {Grindlay}}]{bar00}
{Barret}, D., {Olive}, J.~F., {Boirin}, L., {et~al.} 2000, \apj, 533, 329

\bibitem[{{Belloni} {et~al.}(2005){Belloni}, {Homan}, {Casella}, {van der
  Klis}, {Nespoli}, {Lewin}, {Miller}, \& {Mendez}}]{bel05}
{Belloni}, T., {Homan}, J., {Casella}, P., {et~al.} 2005, astro-ph/0504577

\bibitem[{{Blandford} \& {K\"onigl}(1979)}]{bla79}
{Blandford}, R.~D. \& {K\"onigl}, A. 1979, \apj, 232, 34

\bibitem[{{Buxton} {et~al.}(2012){Buxton}, {Bailyn}, {Capelo}, {Chatterjee},
  {Din{\c c}er}, {Kalemci}, \& {Tomsick}}]{bux12}
{Buxton}, M.~M., {Bailyn}, C.~D., {Capelo}, H.~L., {et~al.} 2012, \aj, 143, 130

\bibitem[{{Cabanac} {et~al.}(2009){Cabanac}, {Fender}, {Dunn}, \&
  {Koerding}}]{cab09}
{Cabanac}, C., {Fender}, R.~P., {Dunn}, R.~J.~H., \& {Koerding}, E.~G. 2009,
  ArXiv e-prints

\bibitem[{{Cadolle Bel} {et~al.}(2011){Cadolle Bel}, {Rodriguez}, {D'Avanzo},
  {Russell}, {Tomsick}, {Corbel}, {Lewis}, {Rahoui}, {Buxton}, {Goldoni}, \&
  {Kuulkers}}]{cad11}
{Cadolle Bel}, M., {Rodriguez}, J., {D'Avanzo}, P., {et~al.} 2011, \aap, 534,
  A119

\bibitem[{{Coppi}(1999)}]{cop99}
{Coppi}, P.~S. 1999, in Astronomical Society of the Pacific Conference Series,
  Vol. 161, High Energy Processes in Accreting Black Holes, ed. J.~{Poutanen}
  \& R.~{Svensson}, 375

\bibitem[{{Corbel} {et~al.}(2013){Corbel}, {Aussel}, {Broderick}, {Chanial},
  {Coriat}, {Maury}, {Buxton}, {Tomsick}, {Tzioumis}, {Markoff}, {Rodriguez},
  {Bailyn}, {Brocksopp}, {Fender}, {Petrucci}, {Cadolle-Bel}, {Calvelo}, \&
  {Harvey-Smith}}]{cor13b}
{Corbel}, S., {Aussel}, H., {Broderick}, J.~W., {et~al.} 2013, \mnras, 431,
  L107

\bibitem[{{Corbel} {et~al.}(2004){Corbel}, {Fender}, {Tomsick}, {Tzioumis}, \&
  {Tingay}}]{cor04}
{Corbel}, S., {Fender}, R.~P., {Tomsick}, J.~A., {Tzioumis}, A.~K., \&
  {Tingay}, S. 2004, \apj, 617, 1272

\bibitem[{{Corbel} {et~al.}(2000){Corbel}, {Fender}, {Tzioumis}, {Nowak},
  {McIntyre}, {Durouchoux}, \& {Sood}}]{cor00}
{Corbel}, S., {Fender}, R.~P., {Tzioumis}, A.~K., {et~al.} 2000, \aap, 359, 251

\bibitem[{{Corbel} {et~al.}(2003){Corbel}, {Nowak}, {Fender}, {Tzioumis}, \&
  {Markoff}}]{cor03}
{Corbel}, S., {Nowak}, M.~A., {Fender}, R.~P., {Tzioumis}, A.~K., \& {Markoff},
  S. 2003, \aap, 400, 1007

\bibitem[{{Coriat} {et~al.}(2011){Coriat}, {Corbel}, {Prat}, {Miller-Jones},
  {Cseh}, {Tzioumis}, {Brocksopp}, {Rodriguez}, {Fender}, \&
  {Sivakoff}}]{cori11}
{Coriat}, M., {Corbel}, S., {Prat}, L., {et~al.} 2011, \mnras, 414, 677

\bibitem[{{Davis} {et~al.}(2006){Davis}, {Done}, \& {Blaes}}]{dav06}
{Davis}, S.~W., {Done}, C., \& {Blaes}, O.~M. 2006, \apj, 647, 525

\bibitem[{{Dhawan} {et~al.}(2000){Dhawan}, {Mirabel}, \&
  {Rodr{\'{\i}}guez}}]{dha00}
{Dhawan}, V., {Mirabel}, I.~F., \& {Rodr{\'{\i}}guez}, L.~F. 2000, \apj, 543,
  373

\bibitem[{{Diaz Trigo} \& {Boirin}(2012)}]{dia12b}
{Diaz Trigo}, M. \& {Boirin}, L. 2012, ArXiv e-prints

\bibitem[{{Diaz Trigo} {et~al.}(2011){Diaz Trigo}, {Miller-Jones}, {Migliari},
  {Parmar}, \& {Boirin}}]{dia11}
{Diaz Trigo}, M., {Miller-Jones}, J., {Migliari}, S., {Parmar}, A., \&
  {Boirin}, L. 2011, in The X-ray Universe 2011, ed. J.-U. {Ness} \& M.~{Ehle},
  3

\bibitem[{{D{\'{\i}}az Trigo} {et~al.}(2012){D{\'{\i}}az Trigo}, {Sidoli},
  {Boirin}, \& {Parmar}}]{dia12}
{D{\'{\i}}az Trigo}, M., {Sidoli}, L., {Boirin}, L., \& {Parmar}, A.~N. 2012,
  \aap, 543, A50

\bibitem[{{Din\c{c}er} {et~al.}(2012){Din\c{c}er}, {Kalemci}, {Buxton},
  {Bailyn}, {Tomsick}, \& {Corbel}}]{din12}
{Din\c{c}er}, T., {Kalemci}, E., {Buxton}, M.~M., {et~al.} 2012, \apj, 753, 55

\bibitem[{{Done} {et~al.}(2012){Done}, {Davis}, {Jin}, {Blaes}, \&
  {Ward}}]{don12}
{Done}, C., {Davis}, S.~W., {Jin}, C., {Blaes}, O., \& {Ward}, M. 2012, \mnras,
  420, 1848

\bibitem[{{Done} \& {Diaz Trigo}(2010)}]{don10}
{Done}, C. \& {Diaz Trigo}, M. 2010, \mnras, 407, 2287

\bibitem[{{Done} {et~al.}(2007){Done}, {Gierlinski}, \& {Kubota}}]{don07}
{Done}, C., {Gierlinski}, M., \& {Kubota}, A. 2007, Astronomy and Astrophysics
  Review, 15, 1

\bibitem[{{Dove} {et~al.}(1997){Dove}, {Wilms}, {Maisack}, \&
  {Begelman}}]{dov97}
{Dove}, J.~B., {Wilms}, J., {Maisack}, M., \& {Begelman}, M.~C. 1997, \apj,
  487, 759

\bibitem[{{Dunn} {et~al.}(2010){Dunn}, {Fender}, {K{\"o}rding}, {Belloni}, \&
  {Cabanac}}]{dun10}
{Dunn}, R.~J.~H., {Fender}, R.~P., {K{\"o}rding}, E.~G., {Belloni}, T., \&
  {Cabanac}, C. 2010, \mnras, 403, 61

\bibitem[{{Esin} {et~al.}(1997){Esin}, {McClintock}, \& {Narayan}}]{esi97}
{Esin}, A.~A., {McClintock}, J.~E., \& {Narayan}, R. 1997, \apj, 489, 865

\bibitem[{{Fabian} {et~al.}(2014){Fabian}, {Parker}, {Wilkins}, {Miller},
  {Kara}, {Reynolds}, \& {Dauser}}]{fab14}
{Fabian}, A.~C., {Parker}, M.~L., {Wilkins}, D.~R., {et~al.} 2014, ArXiv
  e-prints

\bibitem[{{Fender} {et~al.}(1999){Fender}, {Corbel}, {Tzioumis}, {McIntyre},
  {Campbell-Wilson}, {Nowak}, {Sood}, {Hunstead}, {Harmon}, {Durouchoux}, \&
  {Heindl}}]{fen99}
{Fender}, R., {Corbel}, S., {Tzioumis}, T., {et~al.} 1999, \apjl, 519, L165

\bibitem[{{Fender} {et~al.}(2004){Fender}, {Wu}, {Johnston}, {Tzioumis},
  {Jonker}, {Spencer}, \& {van der Klis}}]{fen04}
{Fender}, R., {Wu}, K., {Johnston}, H., {et~al.} 2004, \nat, 427, 222

\bibitem[{{Ferreira} {et~al.}(2006){Ferreira}, {Petrucci}, {Henri}, {Sauge}, \&
  {Pelletier}}]{fer06a}
{Ferreira}, J., {Petrucci}, P.-O., {Henri}, G., {Sauge}, L., \& {Pelletier}, G.
  2006, \aap, 447, 813

\bibitem[{{Gallo} {et~al.}(2003){Gallo}, {Fender}, \& {Pooley}}]{gal03}
{Gallo}, E., {Fender}, R.~P., \& {Pooley}, G.~G. 2003, \mnras, 344, 60

\bibitem[{Gierlinski {et~al.}(1997)Gierlinski, Zdziarski, Done, Johnson,
  Ebisawa, Ueda, Haardt, \& Phlips}]{gier97}
Gierlinski, M., Zdziarski, A.~A., Done, C., {et~al.} 1997, Monthly Notices of
  the Royal Astronomical Society, 288, 958

\bibitem[{{Joinet} {et~al.}(2007){Joinet}, {Jourdain}, {Malzac}, {Roques},
  {Corbel}, {Rodriguez}, \& {Kalemci}}]{joi07}
{Joinet}, A., {Jourdain}, E., {Malzac}, J., {et~al.} 2007, \apj, 657, 400

\bibitem[{{Juett} {et~al.}(2006){Juett}, {Schulz}, {Chakrabarty}, \&
  {Gorczyca}}]{jue06}
{Juett}, A.~M., {Schulz}, N.~S., {Chakrabarty}, D., \& {Gorczyca}, T.~W. 2006,
  \apj, 648, 1066

\bibitem[{{Kalemci} {et~al.}(2013){Kalemci}, {Din{\c c}er}, {Tomsick},
  {Buxton}, {Bailyn}, \& {Chun}}]{kal13}
{Kalemci}, E., {Din{\c c}er}, T., {Tomsick}, J.~A., {et~al.} 2013, \apj, 779,
  95

\bibitem[{{Kong}(2008)}]{kon08}
{Kong}, A.~K.~H. 2008, The Astronomer's Telegram, 1588, 1

\bibitem[{{Kubota} {et~al.}(1998){Kubota}, {Tanaka}, {Makishima}, {Ueda},
  {Dotani}, {Inoue}, \& {Yamaoka}}]{kub98}
{Kubota}, A., {Tanaka}, Y., {Makishima}, K., {et~al.} 1998, \pasj, 50, 667

\bibitem[{{Laor}(1991)}]{lao91}
{Laor}, A. 1991, \apj, 376, 90

\bibitem[{{Luo} \& {Fang}(2014)}]{luo14}
{Luo}, Y. \& {Fang}, T. 2014, \apj, 780, 170

\bibitem[{{Malzac} {et~al.}(2001){Malzac}, {Beloborodov}, \&
  {Poutanen}}]{mal01}
{Malzac}, J., {Beloborodov}, A.~M., \& {Poutanen}, J. 2001, \mnras, 326, 417

\bibitem[{{Merloni} {et~al.}(2000){Merloni}, {Fabian}, \& {Ross}}]{mer00}
{Merloni}, A., {Fabian}, A.~C., \& {Ross}, R.~R. 2000, \mnras, 313, 193

\bibitem[{{Meyer} {et~al.}(2000){Meyer}, {Liu}, \& {Meyer-Hofmeister}}]{mey00a}
{Meyer}, F., {Liu}, B.~F., \& {Meyer-Hofmeister}, E. 2000, \aap, 354, L67

\bibitem[{{Miller} {et~al.}(2002){Miller}, {Ballantyne}, {Fabian}, \&
  {Lewin}}]{mil02a}
{Miller}, J.~M., {Ballantyne}, D.~R., {Fabian}, A.~C., \& {Lewin}, W.~H.~G.
  2002, \mnras, 335, 865

\bibitem[{{Miller} {et~al.}(2006{\natexlab{a}}){Miller}, {Homan}, \&
  {Miniutti}}]{mil06a}
{Miller}, J.~M., {Homan}, J., \& {Miniutti}, G. 2006{\natexlab{a}}, \apjl, 652,
  L113

\bibitem[{{Miller} {et~al.}(2006{\natexlab{b}}){Miller}, {Homan}, {Steeghs},
  {Rupen}, {Hunstead}, {Wijnands}, {Charles}, \& {Fabian}}]{mil06b}
{Miller}, J.~M., {Homan}, J., {Steeghs}, D., {et~al.} 2006{\natexlab{b}}, \apj,
  653, 525

\bibitem[{{Miller} {et~al.}(2006{\natexlab{c}}){Miller}, {Raymond}, {Fabian},
  {Steeghs}, {Homan}, {Reynolds}, {van der Klis}, \& {Wijnands}}]{mil06c}
{Miller}, J.~M., {Raymond}, J., {Fabian}, A., {et~al.} 2006{\natexlab{c}},
  \nat, 441, 953

\bibitem[{{Miller} {et~al.}(2004){Miller}, {Raymond}, {Fabian}, {Homan},
  {Nowak}, {Wijnands}, {van der Klis}, {Belloni}, {Tomsick}, {Smith},
  {Charles}, \& {Lewin}}]{mil04a}
{Miller}, J.~M., {Raymond}, J., {Fabian}, A.~C., {et~al.} 2004, \apj, 601, 450

\bibitem[{{Miller} {et~al.}(2006{\natexlab{d}}){Miller}, {Raymond}, {Homan},
  {Fabian}, {Steeghs}, {Wijnands}, {Rupen}, {Charles}, {van der Klis}, \&
  {Lewin}}]{mil06d}
{Miller}, J.~M., {Raymond}, J., {Homan}, J., {et~al.} 2006{\natexlab{d}}, \apj,
  646, 394

\bibitem[{{Neilsen} \& {Lee}(2009)}]{nei09}
{Neilsen}, J. \& {Lee}, J.~C. 2009, \nat, 458, 481

\bibitem[{{Petrucci} {et~al.}(2008){Petrucci}, {Ferreira}, {Henri}, \&
  {Pelletier}}]{pet08}
{Petrucci}, P.-O., {Ferreira}, J., {Henri}, G., \& {Pelletier}, G. 2008,
  \mnras, 385, L88

\bibitem[{{Plant} {et~al.}(2013){Plant}, {Fender}, {Ponti}, {Munoz-Darias}, \&
  {Coriat}}]{pla13}
{Plant}, D.~S., {Fender}, R.~P., {Ponti}, G., {Munoz-Darias}, T., \& {Coriat},
  M. 2013, ArXiv e-prints

\bibitem[{{Ponti} {et~al.}(2012){Ponti}, {Fender}, {Begelman}, {Dunn},
  {Neilsen}, \& {Coriat}}]{pon12}
{Ponti}, G., {Fender}, R.~P., {Begelman}, M.~C., {et~al.} 2012, \mnras, L417

\bibitem[{{Poutanen} {et~al.}(1997){Poutanen}, {Krolik}, \& {Ryde}}]{pou97}
{Poutanen}, J., {Krolik}, J.~H., \& {Ryde}, F. 1997, \mnras, 292, L21

\bibitem[{{Protassov} {et~al.}(2002){Protassov}, {van Dyk}, {Connors},
  {Kashyap}, \& {Siemiginowska}}]{pro02}
{Protassov}, R., {van Dyk}, D.~A., {Connors}, A., {Kashyap}, V.~L., \&
  {Siemiginowska}, A. 2002, \apj, 571, 545

\bibitem[{{Reis} {et~al.}(2008){Reis}, {Fabian}, {Ross}, {Miniutti}, {Miller},
  \& {Reynolds}}]{rei08}
{Reis}, R.~C., {Fabian}, A.~C., {Ross}, R.~R., {et~al.} 2008, \mnras, 387, 1489

\bibitem[{{Remillard} \& {McClintock}(2006)}]{rem06}
{Remillard}, R.~A. \& {McClintock}, J.~E. 2006, \araa, 44, 49

\bibitem[{{Reynolds} \& {Miller}(2013)}]{rey13}
{Reynolds}, M.~T. \& {Miller}, J.~M. 2013, \apj, 769, 16

\bibitem[{{Ross} \& {Fabian}(2005)}]{ros05}
{Ross}, R.~R. \& {Fabian}, A.~C. 2005, \mnras, 358, 211

\bibitem[{{Russell} {et~al.}(2008){Russell}, {Altamirano}, {Lewis}, {Roche},
  {Markwardt}, \& {Fender}}]{rus08b}
{Russell}, D.~M., {Altamirano}, D., {Lewis}, F., {et~al.} 2008, The
  Astronomer's Telegram, 1586, 1

\bibitem[{{Rykoff} {et~al.}(2007){Rykoff}, {Miller}, {Steeghs}, \&
  {Torres}}]{ryk07}
{Rykoff}, E.~S., {Miller}, J.~M., {Steeghs}, D., \& {Torres}, M.~A.~P. 2007,
  \apj, 666, 1129

\bibitem[{{Salvesen} {et~al.}(2013){Salvesen}, {Miller}, {Reis}, \&
  {Begelman}}]{sal13}
{Salvesen}, G., {Miller}, J.~M., {Reis}, R.~C., \& {Begelman}, M.~C. 2013,
  \mnras

\bibitem[{{Shimura} \& {Takahara}(1995)}]{shi95}
{Shimura}, T. \& {Takahara}, F. 1995, \apj, 445, 780

\bibitem[{{Sobczak} {et~al.}(1999){Sobczak}, {McClintock}, {Remillard},
  {Levine}, {Morgan}, {Bailyn}, \& {Orosz}}]{sob99}
{Sobczak}, G.~J., {McClintock}, J.~E., {Remillard}, R.~A., {et~al.} 1999,
  \apjl, 517, L121

\bibitem[{{Sobolewska} {et~al.}(2011){Sobolewska}, {Papadakis}, {Done}, \&
  {Malzac}}]{sob11}
{Sobolewska}, M.~A., {Papadakis}, I.~E., {Done}, C., \& {Malzac}, J. 2011,
  \mnras, 417, 280

\bibitem[{Stirling {et~al.}(2001)Stirling, Spencer, de~la Force, Garrett,
  Fender, \& Ogley}]{stir01}
Stirling, A.~M., Spencer, R.~E., de~la Force, C.~J., {et~al.} 2001, Monthly
  Notices of the Royal Astronomical Society, 327, 1273, (c) 2001 The Royal
  Astronomical Society

\bibitem[{{Tomsick} {et~al.}(2009){Tomsick}, {Yamaoka}, {Corbel}, {Kaaret},
  {Kalemci}, \& {Migliari}}]{tom09}
{Tomsick}, J.~A., {Yamaoka}, K., {Corbel}, S., {et~al.} 2009, \apjl, 707, L87

\bibitem[{{Wilms} {et~al.}(2000){Wilms}, {Allen}, \& {McCray}}]{wil00}
{Wilms}, J., {Allen}, A., \& {McCray}, R. 2000, \apj, 542, 914

\bibitem[{{Zdziarski} {et~al.}(2004){Zdziarski}, {Gierlinski}, {Miko?ajewska},
  {Wardzinski}, {Smith}, {Alan Harmon}, \& {Kitamoto}}]{zdz04b}
{Zdziarski}, A.~A., {Gierlinski}, M., {Miko?ajewska}, J., {et~al.} 2004,
  \mnras, 351, 791

\bibitem[{{Zimmerman} {et~al.}(2005){Zimmerman}, {Narayan}, {McClintock}, \&
  {Miller}}]{zim05}
{Zimmerman}, E.~R., {Narayan}, R., {McClintock}, J.~E., \& {Miller}, J.~M.
  2005, \apj, 618, 832

\bibitem[{{Zycki} {et~al.}(1998){Zycki}, {Done}, \& {Smith}}]{zyc98}
{Zycki}, P.~T., {Done}, C., \& {Smith}, D.~A. 1998, \apjl, 496, L25

\end{thebibliography}

\end{document}